\documentclass[fleqn,usenatbib]{mnras}

\usepackage{newtxtext,newtxmath}

\usepackage[T1]{fontenc}

\DeclareRobustCommand{\VAN}[3]{#2}
\let\VANthebibliography\thebibliography
\def\thebibliography{\DeclareRobustCommand{\VAN}[3]{##3}\VANthebibliography}

\providecommand{\sorthelp}[1]{}

\usepackage{graphicx}	
\usepackage{amsmath}	
\usepackage{mathtools}

\usepackage{dcolumn}
\usepackage{afterpage}
\usepackage{float}
\usepackage{hyperref}
\usepackage{threeparttable}

\title[The Integrated Radio Spectrum of M\,31]{The C-Band All-Sky Survey (C-BASS): New Constraints on the Integrated Radio Spectrum of M\,31}

\author[Stuart E. Harper et al.]{Stuart E. Harper,$\!^{1}$\thanks{E-mail: \url{stuart.harper@manchester.ac.uk}}
Adam~Barr,$\!^{1}$
C.~Dickinson,$\!^{1,2}$\thanks{E-mail: \url{clive.dickinson@manchester.ac.uk}}
M.\,W.~Peel,$\!^{3,4,5}$ 
Roke~Cepeda-Arroita,$\!^{1}$
C.\,J.~Copley,$\!^{6}$ \newauthor
R.\,D.\,P.~Grumitt,$\!^{7}$ 
J.\,Patrick~Leahy,$\!^{1}$ 
J.\,L.~Jonas,$\!^{8,9}$
Michael\,E.~Jones,$\!^7$  
J.~Leech,$\!^{7}$
T.\,J.~Pearson,$\!^{2}$\newauthor
A.\,C.\,S.~Readhead,$\!^{2}$ 
Angela\,C.~Taylor$^7$
  \\
$^1$Jodrell Bank Centre for Astrophysics, Department of Physics \& Astronomy, The University of Manchester, \\
Oxford Road, Manchester, M13 9PL, U.K. \\ 
$^{2}$Cahill Centre for Astronomy and Astrophysics, California Institute of Technology, Pasadena, CA 91125, U.S.A. \\
$^{3}$Instituto de Astrof\'{i}sica de Canarias, E-38205 La Laguna, Tenerife, Spain\\
$^{4}$Departamento de Astrof\'{i}sica, Universidad de La Laguna (ULL), E-38206 La Laguna, Tenerife, Spain\\
$^5$Imperial College London, Blackett Lab, Prince Consort Road, London SW7 2AZ, UK\\
$^6$Ancera.com, 15 Commercial St, Branford, CT 06405, USA \\
$^7$Sub-department of Astrophysics, University of Oxford, Denys Wilkinson Building, Keble Road, Oxford OX1 3RH, U.K. \\
$^{8}$Department of Physics and Electronics, Rhodes University, Grahamstown, 6139, South Africa \\
$^{9}$South African Radio Astronomy Observatory, 2 Fir Road, Observatory, Cape Town, 7925, South Africa\\
}

\date{Accepted XXX. Received YYY; in original form ZZZ}

\pubyear{2022}

\begin{document}

\newcommand{\emissivity}{$0.17 \pm 0.07$\,K/$\tau_{353}$}
\newcommand{\sourcestotal}{332}
\newcommand{\sourcesfitted}{71}
\newcommand{\noCMBsubASync}{$8.3\pm0.8$}
\newcommand{\CMBsubASync}{$9.7\pm0.6$}
\newcommand{\CMBSRCsubASync}{$10.6\pm0.5$}
\newcommand{\CMBSRCsubnoamecurvedASync}{$9.7\pm0.6$}
\newcommand{\CMBSRCsubcurvedASync}{$10.2\pm0.6$}
\newcommand{\CMBSRCsubnoameASync}{$9.9\pm0.5$}
\newcommand{\planckmodelASync}{$9.5\pm1.1$}
\newcommand{\srtmodelASync}{$7.0\pm0.5$}
\newcommand{\noCMBsubAlpha}{$-0.73\pm0.10$}
\newcommand{\CMBsubAlpha}{$-0.63\pm0.05$}
\newcommand{\CMBSRCsubAlpha}{$-0.66\pm0.03$}
\newcommand{\CMBSRCsubnoamecurvedAlpha}{$-0.61\pm0.06$}
\newcommand{\CMBSRCsubcurvedAlpha}{$-0.60\pm0.06$}
\newcommand{\CMBSRCsubnoameAlpha}{$-0.61\pm0.04$}
\newcommand{\planckmodelAlpha}{$-0.92\pm0.16$}
\newcommand{\srtmodelAlpha}{$-1.10\pm0.09$}
\newcommand{\noCMBsubCsCurve}{--}
\newcommand{\CMBsubCsCurve}{--}
\newcommand{\CMBSRCsubCsCurve}{--}
\newcommand{\CMBSRCsubnoamecurvedCsCurve}{$-0.007\pm0.017$}
\newcommand{\CMBSRCsubcurvedCsCurve}{$-0.04\pm0.03$}
\newcommand{\CMBSRCsubnoameCsCurve}{--}
\newcommand{\planckmodelCsCurve}{--}
\newcommand{\srtmodelCsCurve}{--}
\newcommand{\noCMBsubEm}{$3.9\pm1.5$}
\newcommand{\CMBsubEm}{$1.1\pm0.8$}
\newcommand{\CMBSRCsubEm}{$0.4\pm0.4$}
\newcommand{\CMBSRCsubnoamecurvedEm}{$1.2\pm0.9$}
\newcommand{\CMBSRCsubcurvedEm}{$1.1\pm0.8$}
\newcommand{\CMBSRCsubnoameEm}{$0.9\pm0.6$}
\newcommand{\planckmodelEm}{$1.8\pm1.3$}
\newcommand{\srtmodelEm}{$1.1\pm0.9$}
\newcommand{\noCMBsubDt}{$0.3\pm1.1$}
\newcommand{\CMBsubDt}{--}
\newcommand{\CMBSRCsubDt}{--}
\newcommand{\CMBSRCsubnoamecurvedDt}{--}
\newcommand{\CMBSRCsubcurvedDt}{--}
\newcommand{\CMBSRCsubnoameDt}{--}
\newcommand{\planckmodelDt}{--}
\newcommand{\srtmodelDt}{--}
\newcommand{\noCMBsubTD}{$19.5\pm0.8$}
\newcommand{\CMBsubTD}{$17.3\pm0.4$}
\newcommand{\CMBSRCsubTD}{$17.2\pm0.4$}
\newcommand{\CMBSRCsubnoamecurvedTD}{$16.9\pm0.4$}
\newcommand{\CMBSRCsubcurvedTD}{$17.3\pm0.4$}
\newcommand{\CMBSRCsubnoameTD}{$16.9\pm0.4$}
\newcommand{\planckmodelTD}{$18.2\pm1.0$}
\newcommand{\srtmodelTD}{$18.8\pm0.5$}
\newcommand{\noCMBsubTau}{$1.06\pm0.19$}
\newcommand{\CMBsubTau}{$1.78\pm0.18$}
\newcommand{\CMBSRCsubTau}{$1.84\pm0.19$}
\newcommand{\CMBSRCsubnoamecurvedTau}{$2.00\pm0.2$}
\newcommand{\CMBSRCsubcurvedTau}{$1.79\pm0.19$}
\newcommand{\CMBSRCsubnoameTau}{$2.00\pm0.2$}
\newcommand{\planckmodelTau}{$1.2\pm0.2$}
\newcommand{\srtmodelTau}{--}
\newcommand{\noCMBsubBeta}{$1.44\pm0.09$}
\newcommand{\CMBsubBeta}{$1.76\pm0.05$}
\newcommand{\CMBSRCsubBeta}{$1.77\pm0.05$}
\newcommand{\CMBSRCsubnoamecurvedBeta}{$1.81\pm0.05$}
\newcommand{\CMBSRCsubcurvedBeta}{$1.76\pm0.05$}
\newcommand{\CMBSRCsubnoameBeta}{$1.81\pm0.04$}
\newcommand{\planckmodelBeta}{$1.62\pm0.11$}
\newcommand{\srtmodelBeta}{$1.49\pm0.06$}
\newcommand{\noCMBsubSpAAme}{$0.18\pm0.15$}
\newcommand{\CMBsubSpAAme}{$0.32\pm0.11$}
\newcommand{\CMBSRCsubSpAAme}{$0.27\pm0.09$}
\newcommand{\CMBSRCsubnoamecurvedSpAAme}{--}
\newcommand{\CMBSRCsubcurvedSpAAme}{$0.38\pm0.12$}
\newcommand{\CMBSRCsubnoameSpAAme}{--}
\newcommand{\planckmodelSpAAme}{$0.7\pm0.3$}
\newcommand{\srtmodelSpAAme}{$1.45\pm0.18$}
\newcommand{\noCMBsubSpNuAme}{$35\pm13$}
\newcommand{\CMBsubSpNuAme}{$25\pm2$}
\newcommand{\CMBSRCsubSpNuAme}{$24\pm2$}
\newcommand{\CMBSRCsubnoamecurvedSpNuAme}{--}
\newcommand{\CMBSRCsubcurvedSpNuAme}{$26\pm2$}
\newcommand{\CMBSRCsubnoameSpNuAme}{--}
\newcommand{\planckmodelSpNuAme}{--}
\newcommand{\srtmodelSpNuAme}{--}
\newcommand{\noCMBsubChisq}{$9.8$}
\newcommand{\CMBsubChisq}{$20.6$}
\newcommand{\CMBSRCsubChisq}{$18.7$}
\newcommand{\CMBSRCsubnoamecurvedChisq}{$29.2$}
\newcommand{\CMBSRCsubcurvedChisq}{$18.6$}
\newcommand{\CMBSRCsubnoameChisq}{$26.5$}
\newcommand{\planckmodelChisq}{--}
\newcommand{\srtmodelChisq}{--}
\newcommand{\noCMBsubChisqReduced}{$1.4$}
\newcommand{\CMBsubChisqReduced}{$2.6$}
\newcommand{\CMBSRCsubChisqReduced}{$2.3$}
\newcommand{\CMBSRCsubnoamecurvedChisqReduced}{$3.2$}
\newcommand{\CMBSRCsubcurvedChisqReduced}{$2.3$}
\newcommand{\CMBSRCsubnoameChisqReduced}{$2.9$}
\newcommand{\planckmodelChisqReduced}{--}
\newcommand{\srtmodelChisqReduced}{--}
\newcommand{\noCMBsubAicc}{$107$}
\newcommand{\CMBsubAicc}{$110$}
\newcommand{\CMBSRCsubAicc}{$106$}
\newcommand{\CMBSRCsubnoamecurvedAicc}{$124$}
\newcommand{\CMBSRCsubcurvedAicc}{$106$}
\newcommand{\CMBSRCsubnoameAicc}{$119$}
\newcommand{\planckmodelAicc}{--}
\newcommand{\srtmodelAicc}{--}

\newcommand{\ameCurvatureFFFlux}{$S_\mathrm{ff} = 0.33 \pm 0.23$}
\newcommand{\ameCurvatureMurphySFR}{$0.16 \pm 0.12$\,$M_\odot$\,yr$^{-1}$}

\newcommand{\amePlawFFFlux}{$S_\mathrm{ff} = 0.13 \pm 0.13$}
\newcommand{\amePlawMurphySFR}{$< 0.12$\,$M_\odot$\,yr$^{-1}$}

\newcommand{\noamePlawFFFlux}{$S_\mathrm{ff} = 0.25 \pm 0.18$}
\newcommand{\noamePlawMurphySFR}{$0.12 \pm 0.09$\,$M_\odot$\,yr$^{-1}$}

\newcommand{\cbassFluxDensity}{$4.04\pm 0.14$}

\label{firstpage}
\pagerange{\pageref{firstpage}--\pageref{lastpage}}
\maketitle

\begin{abstract}
The Andromeda galaxy (M\,31) is our closest neighbouring spiral galaxy, making it an ideal target for studying the physics of the interstellar medium in a galaxy very similar to our own. Using new observations of M\,31 at 4.76\,GHz by the C-Band All-Sky Survey (C-BASS), and all available radio data at $1^\circ$ resolution, we produce the integrated spectrum and put new constraints on the synchrotron spectral index and anomalous microwave emission (AME) from M\,31. We use aperture photometry and spectral modelling to fit for the integrated spectrum of M\,31, and subtract a comprehensive model of nearby background radio sources. The AME in M\,31 is detected at $3\sigma$ significance with a peak near 30\,GHz and flux density \CMBSRCsubSpAAme\,Jy. The synchrotron spectral index of M\,31 is flatter than our own Galaxy at $\alpha =$\CMBSRCsubAlpha~with no strong evidence of spectral curvature. The emissivity of AME, averaged over the total emission from M\,31 is lower than typical AME sources in our Galaxy, implying that AME is not uniformly distributed throughout M\,31 and instead is likely confined to sub-regions---this will need to be confirmed using future higher resolution observations around 20--30\,GHz. 
\end{abstract}




\section{Introduction}

The Andromeda Galaxy, also known as Messier 31 (M\,31), is the closest spiral galaxy to our Galaxy, its radio continuum extent subtending $\approx2.7^{\circ}\times1^{\circ}$ on the sky \citep{berkhuijsen1983}. The proximity of M\,31 and its similarities to our own Galaxy have made it an ideal target for testing our understanding of the interstellar medium (ISM) and it has been studied at many wavelengths from radio \citep[e.g.,][]{brown2015,Beck1980,berkhuijsen2003}, to IR/sub-mm \citep[e.g.,][]{Gordon2006,fritz2012,planck_andromeda}, optical \citep[e.g.,][]{lauer1993} and UV \citep[e.g.,][]{code1979}. Together these observations have greatly improved our understanding of a range of physical phenomena within M\,31, from the synchrotron emission emitted by cosmic rays, to thermal dust properties, and the physics of star-formation within a galaxy like our own.

At radio frequencies ($\lesssim 100$\,GHz) the emission from normal galaxies is typically due to synchrotron and free-free processes, while at higher frequencies ($\gtrsim 100$\,GHz) thermal dust radiation dominates. Anomalous microwave emission (AME), first observed several decades ago \citep{kogut1996,leitch1997}, is a fourth component that is characterized by a spectrum that is peaked around 30 GHz. The most favoured model for AME is that it is generated by rapidly rotating spinning dust grains \citep{draine1998b,ali-hamoud2009}. AME has been found in a range of Galactic structures from high Galactic latitude cirrus clouds \citep[e.g.][]{Davies2006,planck2013-p06,Harper2022}, to individual dust clouds and \textsc{Hii} regions \citep{Watson2005,Casassus2008,Cepeda-Arroita2021,Rennie2022}. Yet, AME in other galaxies seems to be far less prevalent with little or no evidence of AME in the integrated spectra of several spiral galaxies \citep{peel_2011,Bianchi2022}, and the only clear detection of AME in external galaxies being confined to individual star-forming regions \citep{Hensley2015,Murphy2018}. When estimating the average emissivity of the AME (i.e., the AME brightness per unit dust column) from other galaxies we find that it is much lower than known sources of AME implying that AME is generally not uniformly distributed throughout the ISM. A detailed review of current AME science can be found in \citet{dickinson2018}. 

The first attempt to characterise AME in M\,31 used a combination of WMAP and \textit{Planck} data alongside older radio surveys \citep{planck_andromeda}, and found there is a tentative evidence for AME in the integrated spectrum of M\,31. Later, observations at $\approx 6$\,GHz by the Sardinian Radio Telescope (SRT) completely changed the prospect of AME in M\,31, suggesting that AME is the dominant emission component at 30\,GHz and the emissivity was similar to that observed in our own galaxy \citep{battistelli_andromeda}---something not found in other galaxies, or by the \cite{planck_andromeda} analysis. 

We revisit the integrated spectrum of M\,31 using observations from the C-Band All Sky Survey (C-BASS) at 4.76\,GHz (Taylor et al. in prep.), and the publicly available QUI Joint Tenerife Experiment (QUIJOTE) wide survey \citep{RubinoMartin2023} observations at 11 to 19\,GHz. These two surveys fill in the gap between WMAP/\textit{Planck} microwave frequencies and lower frequency radio data. Unlike the SRT data, the C-BASS and QUIJOTE observations have resolutions comparable to the other surveys, making them ideal for constraining the low frequency spectrum of M\,31. 

The paper is structured as follows: In \autoref{sec:data} we give an overview of the \mbox{C-BASS} experiment and the ancillary datasets used. In \autoref{sec:photometry} we give an overview of how we perform aperture photometry and the model fitted to the SED of M\,31. \autoref{sec:sources} describes how we model and remove other extragalactic radio sources from the region around M\,31. In \autoref{sec:results} we present the main results, followed by a discussion in \autoref{sec:discussion}. Finally, \autoref{sec:conclusions} summarises the main conclusions.

\section{Data}
\label{sec:data}

\subsection{C-BASS}
\label{sec:cbass}

The C-Band All Sky Survey (\mbox{C-BASS}) is a ground-based full-sky survey of total intensity and polarisation at a frequency of 4.76\,GHz~\citep{Jones2018}. The survey has an angular resolution of $0.\!\!^\circ73$ full-width half-maximum (FWHM). \mbox{C-BASS} observed the northern hemisphere from the Owens Valley Radio Observatory in California, USA, and observations of the southern sky are ongoing at Klerefontein, South Africa. In this work we use the \mbox{C-BASS} North total intensity data that has been deconvolved to a Gaussian beam with a FWHM of $1^\circ$. Below we give a brief summary of the \mbox{C-BASS} instrument, data processing, and maps; full details will be published in \mbox{Taylor et al. (in prep.)}.

The \mbox{C-BASS} North telescope is a 6.1\,m Gregorian antenna. To suppress sidelobe pick-up and symmetrise the beam response the primary antenna was under-illuminated; a radio-absorbing baffle surrounded the edge of the primary; and the struts supporting the secondary were replaced with a low-loss dielectric cone \citep{Holler2013}. The telescope beam contains 73\,per\,cent of the power in the main beam, and 95\,per\,cent within $9.\!\!^\circ2$ radius.

The receiver was a cryogenically cooled dual circularly polarised correlation receiver that allows for the simultaneous measurement of Stokes I, Q, and U. The receiver has a nominal bandwidth of 4.5--5.5\,GHz, but the effective bandwidth was 0.5\,GHz due to notch filters installed around the centre of the band to suppress fixed terrestrial radio frequency interference (RFI) \citep{King2014}.

\mbox{C-BASS} North observations were taken between 2012 November and 2015 March, with a total observing time of approximately 2000\,hours. The observing strategy was to perform $360^\circ$ azimuth sweeps at fixed elevations. The total sky area covered was 26000\,sq.\,degrees and the minimum declination observed was $-15^\circ$.

The \mbox{C-BASS} north data processing pipeline will be described in detail in Taylor et al. (in prep.). The data processing includes masking transient RFI and solar system objects, modelling and subtracting ground emission, and suppressing a microphonic 1.2\,Hz oscillation in the time-ordered data (TOD) that is due to the receiver cryogenics.

The \mbox{C-BASS} maps are made using the \texttt{Descart} \citep{Sutton2010} implementation of the destriping map-making method. Destriping map-making fits offsets to the time correlated $1/f$ noise in the \mbox{C-BASS} TOD by using the covariance of the sky signal and \mbox{C-BASS} observing strategy. Jack-knife tests of the \mbox{C-BASS} data found that residual $1/f$ noise and other systematics after map-making are less than a 1\,per\,cent on scales of the \mbox{C-BASS} beam.

The absolute temperature scale of the \mbox{C-BASS} map was set by the WMAP models of the flux density and secular changes of the supernova remnant Taurus\,A (Tau\,A) \citep{Weiland2011}. The flux density models for such bright calibrator sources (e.g., Cas\,A and Tau\,A) are known to $\approx 0.5$ per cent precision, providing a precise absolute calibration scale. The band-averaged central frequency of \mbox{C-BASS} when calibrated against a flat-spectrum source ($\beta = -2$ in brightness temperature units) was 4.76\,GHz. The calibration uncertainty of the \mbox{C-BASS} intensity map has additional contributions from colour corrections, and the beam model. The total calibration uncertainty of the \mbox{C-BASS} North map is conservatively assumed to be 3\,per\,cent.

\begin{table*}
        \centering
        \caption{The integrated flux densities of M\,31 for the three cases considered: without subtracting the CMB, subtracting the SMICA CMB model, and lastly  subtracting our model of background radio sources at each frequency as well as the SMICA CMB model. We include the flux densities from the \textit{Planck} \citep{planck_andromeda} and SRT analysis \citep{battistelli_andromeda} for comparison.}
        \begin{tabular}{cccccccc}
        \hline
               &           &          & No CMB      & CMB        & CMB \& Source  &                              &   \\
        Survey & Frequency & Cal.\!$^4$ & subtraction & Subtracted & Subtracted     & \citet{planck_andromeda}$^2$ & \citet{battistelli_andromeda}$^3$ \\
         & (GHz) & (\%) & (Jy) & (Jy) & (Jy) & (Jy) \\
        \hline
        \hline
Haslam & 0.408 & 10.0 & 17.5$\pm$1.9 & 17.5$\pm$1.9 & 18$\pm$2 & 23$\pm$4 & 18.4$\pm$1.6 \\
SRT HI-1 & 1.385 & 5.0 & -- & -- & -- & -- & 5.4$\pm$0.4\\
Reich & 1.42 & 10.0 & 8.5$\pm$0.7 & 8.5$\pm$0.7 & 8.4$\pm$0.7 & -- & -- \\
SRT HI-2 & 1.437 & 5.0 & -- & -- & -- & -- & 5.3$\pm$0.4\\
C-BASS & 4.76 & 3.0 & 4.06$\pm$0.14 & 4.06$\pm$0.14 & 4.04$\pm$0.14 & -- & -- \\
SRT C-1 & 6.313 & 5.0 & -- & -- & -- & -- & 1.21$\pm$0.08\\
SRT C-2 & 6.938 & 5.0 & -- & -- & -- & -- & 1.19$\pm$0.09\\
QUIJOTE & 12.0 & 5.0 & 2.73$\pm$0.16 & 2.75$\pm$0.16 & 2.48$\pm$0.16 & -- & -- \\
WMAP & 22.8 & 3.0 & 2.27$\pm$0.11 & 2.12$\pm$0.08 & 1.84$\pm$0.07 & 2.09$\pm$0.10 & 2.00$\pm$0.17 \\
$\mathit{Planck}$ & 28.4 & 1.0 & 2.28$\pm$0.11 & 1.95$\pm$0.05 & 1.71$\pm$0.06 & 2.05$\pm$0.11 & 1.86$\pm$0.15 \\
WMAP & 33.0 & 3.0 & 2.11$\pm$0.15 & 1.85$\pm$0.08 & 1.54$\pm$0.08 & 1.88$\pm$0.14 & 1.71$\pm$0.21 \\
WMAP & 40.7 & 3.0 & 2.00$\pm$0.20 & 1.48$\pm$0.10 & 1.22$\pm$0.10 & 1.73$\pm$0.18 & 1.31$\pm$0.16 \\
$\mathit{Planck}$ & 44.1 & 1.0 & 2.07$\pm$0.24 & 1.53$\pm$0.11 & 1.24$\pm$0.11 & 1.31$\pm$0.21 & 1.45$\pm$0.25 \\
WMAP & 60.7 & 3.0 & 3.6$\pm$0.4 & 2.27$\pm$0.20 & 1.90$\pm$0.20 & 2.9$\pm$0.4 & 1.7$\pm$0.4 \\
$\mathit{Planck}$ & 70.4 & 1.0 & 3.5$\pm$0.6 & 1.83$\pm$0.24 & 1.51$\pm$0.24 & 3.3$\pm$0.5 & 2.1$\pm$0.4 \\
WMAP & 93.5 & 3.0 & 5.1$\pm$0.9 & 2.9$\pm$0.4 & 2.6$\pm$0.4 & 4.3$\pm$0.8 & 3.5$\pm$1.0 \\
$\mathit{Planck}$ & 100.0 & 1.0 & 8.2$\pm$1.0 & 5.3$\pm$0.4 & 5.0$\pm$0.4 & 7.3$\pm$1.2 & 5.8$\pm$0.5 \\
$\mathit{Planck}$ & 143.0 & 1.0 & 20.4$\pm$1.6 & 14.2$\pm$0.7 & 14.0$\pm$0.7 & 18.2$\pm$1.5 & 15.7$\pm$1.4 \\
$\mathit{Planck}$ & 217.0 & 1.0 & 80$\pm$3 & 68.2$\pm$1.5 & 67.9$\pm$1.5 & 76$\pm$8 & 69$\pm$6 \\
$\mathit{Planck}$ & 353.0 & 1.3 & 325$\pm$7 & 318$\pm$6 & 318$\pm$6 & 298$\pm$13 & 318$\pm$24 \\
$\mathit{Planck}$ & 545.0 & 6.0 & 1054$\pm$65 & 1054$\pm$65 & 1054$\pm$65 & 1020$\pm$100 & 1027$\pm$73 \\
$\mathit{Planck}$ & 857.0 & 6.4 & 3120$\pm$210 & 3120$\pm$210 & 3120$\pm$210 & 3050$\pm$310 & 3020$\pm$190 \\
COBE-DIRBE & 1249.0 & 13.5 & 5890$\pm$810 & 5890$\pm$810 & 5890$\pm$810 & 5700$\pm$770 & -- \\
COBE-DIRBE & 2141.0 & 10.6 & 7310$\pm$810 & 7310$\pm$810 & 7310$\pm$810 & 7300$\pm$1000 & -- \\
COBE-DIRBE & 3000.0 & 11.6 & 3570$\pm$440 & 3570$\pm$440 & 3570$\pm$440 & 3600$\pm$500 & -- \\
\hline

\hline
    \end{tabular}
\begin{tablenotes}
\item $^1$ We combine both the 11 and 13\,GHz data into a single channel, see \autoref{sec:ancillary} for details. 
\item $^2$ The \textit{Planck} analysis includes contributions from the CMB and background sources.
\item $^3$ The SRT analysis subtracted the a CMB model and background radio sources.
\item $^4$ Calibration uncertainties for each survey are discussed in \autoref{sec:cbass} and \autoref{sec:ancillary}.
\end{tablenotes}
\label{tab:fluxes}
\end{table*}

\subsection{Ancillary Data}
\label{sec:ancillary}

We use additional surveys at radio and infrared bands to fully sample the spectrum of M\,31 around the peak in the spinning dust emission. A summary of these ancillary data is given in \autoref{tab:fluxes}. All maps are smoothed to a common $1^{\circ}$ resolution and repixelized into {\tt HEALPix} $N_{\mathrm{side}}=256$ maps.

The well-known Haslam 408\,MHz map is used to constrain low-frequency emission~\citep{haslam1,haslam2}. We use the  version of the 408\,MHz map that was reprocessed by \citet{remazeilles2015}, which significantly reduces the striping artefacts due to $1/f$ noise along the scan directions. The survey has a native resolution of approximately 56\,arcmin FWHM. The 408\,MHz survey is calibrated against an absolutely calibrated 404\,MHz survey \citep{Pauliny-Toth1962} with a nominal calibration uncertainty of 5\,per\,cent. However, uncertainties in the beam of the 408\,MHz are not well known and can translate into scale-dependent changes in the flux density across the map that could be up to tens of percent. There are also residual $1/f$ noise stripes that can contribute up to 3.8\,Jy per beam. As such, to account for all these effects we conservatively assign a 10\,per\,cent calibration uncertainty to the 408\,MHz data, similar to other analyses \citep[e.g.,][]{planck_ame}.

The full-sky map at 1.42\,GHz~\citep{reich86,reich1988} combines data from the German Stockert 25-m and Argentinian Villa Elisa 30-m telescopes to produce a full-sky map with 36\,arcmin resolution. We use a destriped version of the map, calibrated using absolute sky horn measurements. It is common to apply a main-beam to full-beam correction factor of 1.55 to bring the absolute calibration onto the main-beam scale, however \citet{planck_andromeda} suggests a factor of 1.3 would be better as the source is partially resolved. We adopt the factor of 1.3 for this analysis, but the results do not depend significantly on which factor is chosen. As with the 408\,MHz data there are large uncertainties in this map related to the scale-dependent changes in flux density due to the beam, residual noise stripes, and the inherent calibration uncertainty. \citet{reich1988} give an uncertainty on the full-beam to main-beam ratio of approximately 5\,per\,cent when directly comparing the flux density conversion factors of the Stockert data to that of an older radio survey, however when doing a comparison of these two surveys at a common resolution they find a 16\,per\,cent scatter in the measured pixel brightnesses. Noting that these uncertainties include contributions from two surveys, and that there is no simple way to determine precisely full-beam to main-beam ratio for a resolved source like M\,31 in the Stockert data, we assign the 1.42\,GHz survey a conservative 10\,per\,cent calibration uncertainty.

The QUI Joint Tenerife Experiment (QUIJOTE) Multi-Frequency Instrument (MFI) is a four horn focal plane array that observes between 10 and 20\,GHz at $1^\circ$ resolution \citep{RubinoMartin2023}. The QUIJOTE MFI wide survey data covers the entire Northern sky including M\,31. We use the 2023 data release available on the NASA LAMBDA website\footnote{\url{https://lambda.gsfc.nasa.gov/product/quijote/index.html}}. For this analysis we combine the MFI 11 and 13\,GHz channels together and then smooth the combined map to $1^\circ$ resolution assuming a weighted average of the FWHM for both the 11 and 13\,GHz channels of $55.\!\!^{\circ}61$ and a frequency of 12\,GHz. The reason for combining the maps is the high correlation in the noise between the two channels, which is  90--95\,per\,cent in intensity \citep{RubinoMartin2023}. The 17 and 19\,GHz channels do not have sufficient sensitivity to detect M\,31 and are therefore excluded. The calibration uncertainty for the QUIJOTE MFI data is 5\,per\,cent. 

We use maps from the \textit{Planck} \texttt{NPIPE} joint LFI and HFI reprocessing \citep{planck_npipe} with eight frequency bands ranging from 28.4\,GHz to 857\,GHz. It should be noted that this is different from the~\citet{planck_andromeda} analysis, which uses the maps from the 2015 data release. We do not use the 100\,GHz and 217\,GHz maps, as these frequency bands contain contamination from CO molecular lines~\citep[e.g.,][]{planck_ame}. This has a significant effect on the estimate of the flux densities at these frequencies, which in turn affects the modelling of the thermal dust emission. We adopt calibration uncertainties for the LFI of 3\,per\,cent \citep{planck_ame} and HFI of 5\,per\,cent  \citep{planck2018}; these uncertainties account for colour corrections, residual beam asymmetries, and other low-level systematic errors.

WMAP has five frequency bands: 22.8\,GHz, 33.0\,GHz, 40.7\,GHz, 60.7\,GHz and 93.5\,GHz. The maps, smoothed to 1$^\circ$, are the final 9-year maps \citep{bennett2013} stored in the LAMBDA\footnote{\url{https://lambda.gsfc.nasa.gov}} archive. A 3\,per\,cent overall calibration uncertainty was applied to take into account potential low-level systematics such as beam asymmetries, which was used by other studies \citep{planck_ame}. We believe this is reasonable since we are not limited by instrumental noise.

Infrared surveys let us measure the (modified) blackbody curve of thermal dust at temperatures above $\approx 15$\,K. The COBE-DIRBE survey \citep{hauser1998}, at 240\,$\mu$m (1249\,GHz) and 140\,$\mu$m (2141\,GHz) are used to measure the peak of the thermal dust spectrum. The Zodi-Subtracted Mission Average (ZSMA) maps are used, regridded to the {\tt HEALPix} format.

\section{Photometry and Modelling}
\label{sec:photometry}

 \subsection{Spectral models}
 \label{sec:models}

 The integrated emission from M\,31 is modelled as a combination of up to five components: synchrotron, free-free, spinning dust, thermal dust emission, and the CMB. The full model is defined as
 \begin{flalign}
     S_{\mathrm{total}}(\nu)= ~& S_{\mathrm{synch}}(\nu,A_{\mathrm{synch}},\alpha) \nonumber
     \\ \nonumber &+ S_{\mathrm{ff}}(\nu,\mathrm{EM})+S_{\mathrm{AME}}(\nu, A_{\mathrm{AME}},\nu_{\mathrm{AME}}) 
     \\  &+S_{\mathrm{td}}(\nu,\tau_{250},T_{\mathrm{td}}, \beta)  +S_\mathrm{CMB}(\nu, \Delta T),
\end{flalign}
 where each contribution is explained in detail below.

 The synchrotron emission is modelled as a simple power-law, with two free parameters:
 \begin{equation}
     S_{\mathrm{synch}}(\nu)= A_{\mathrm{synch}} \nu^{\alpha},
 \end{equation}
 where $A_{\mathrm{synch}}$ is the synchrotron amplitude at 1\,GHz, $\nu$ is the frequency in GHz and $\alpha$ is the flux density spectral index. We also include an alternative synchrotron model that allows for spectral curvature:
  \begin{equation}\label{eqn:powerlaw_curvature}
     S_{\mathrm{synch}}(\nu)= A_{\mathrm{synch}} \nu^{\alpha + C \log(\nu)},
 \end{equation}
 where $C$ is the spectral curvature parameter. 

 The free-free flux density is converted from a free-free brightness temperature, $T_{\mathrm{ff}}$, by
 \begin{equation}
 S_{\mathrm{ff}}=\frac{2kT_{\mathrm{ff}}\Omega\nu^2}{c^2},
 \end{equation}
 where $k$ is the Boltzmann constant, $\Omega$ the solid angle of the aperture, and $\nu$ is the observing frequency. The free-free brightness temperature is given by
 \begin{equation}
 T_{\mathrm{ff}}= T_{\mathrm{e}}\left (1-e^{-\tau_{\mathrm{ff}}} \right ),
 \end{equation}
 where $T_{\mathrm{e}}$ is the electron temperature and $\tau_{\mathrm{ff}}$ is the free-free optical depth. The optical depth $\tau_{\mathrm{ff}}$ is given by \citet{draine2011}
 \begin{equation}
     \tau_{\mathrm{ff}}= 5.468\times10^{-2}T_{\mathrm{e}}^{-1.5}\nu_\mathrm{GHz}^{-2} \mathrm{EM} \, g_{\mathrm{ff}},
 \end{equation}
 where $\mathrm{EM}$ (pc\,cm$^{-6}$) is the emission measure and $g_{\mathrm{ff}}$ is the Gaunt factor. We use the approximation for the Gaunt factor derived by \citet{draine2011}:
 \begin{equation}
 g_{\mathrm{ff}}\approx \ln \left \{ \mathrm{exp}\left [ 5.960-\frac{\sqrt{3}}{\pi}\ln\left ( Z_i\nu_{\mathrm{GHz}}T_4^{-3/2}\right ) \right ] + e\right \},
 \end{equation}
 where $e$ is Euler's constant, $\nu_{\mathrm{GHz}}$ is the frequency in GHz, $T_4$ is the temperature in units of $10^4$\,K and $Z_i$ is the charge on the scattering ions (assumed to be singly ionised). We use a fixed electron temperature of 8000\,K, which is typical for most Galactic sources \citep{Paladini2003}, and only fit for $\mathrm{EM}$. Such assumptions have negligible impact on the spectral model.

 The thermal dust is modelled as a modified blackbody. It is fitted using \begin{equation}
     S_{\mathrm{td}}= 2 h \frac{\nu^3}{c^2} \frac{1}{e^{h\nu/kT_\mathrm{d}}-1}\tau_{250}(\nu/1.2 \mathrm{THz})^{\beta_\mathrm{d}}\Omega,
 \end{equation}
 where $\tau_{250}$ is the optical depth at 250$\mu$m, $T_\mathrm{d}$ is the dust temperature, and $\beta_\mathrm{d}$ is the dust emissivity spectral index. Using a single-component modified blackbody model is a simplification of the thermal dust emission spectrum since we know in M\,31 there will be many dust components with different emissivities, and temperatures. However, this approximation will only cause issues around the peak of the thermal dust emission spectrum at $\nu > 1000$\,GHz, which is not critical for characterising the AME and the spectrum at lower frequencies.


 We model the spinning dust component as a fixed template, produced using the \textsc{SpDust} code~\citep{ali-hamoud2009}. The chosen template represents the warm neutral medium (WNM) as defined by \cite{draine1998b}, which is generally representative of AME regions \citep{dickinson2018}. This template has a peak frequency of 28.1\,GHz. The overall shape of the spinning dust spectrum is not strongly dependent on the environment in this analysis as the choice of environment mostly affects the spinning dust peak frequency and amplitude both of which are free parameters in this analysis. To fit the spinning dust component to an arbitrary peak frequency we interpolate the WNM model from \textsc{SpDust} such that
 \begin{equation}
     S_{\mathrm{AME}}=A_{\mathrm{AME}}\frac{j(\nu \times \nu_\mathrm{WNM}/\nu_0)}{j(\nu_\mathrm{r}\times \nu_\mathrm{WNM}/\nu_0)}\Omega,
 \end{equation}
where $\nu$ is the input frequency, $\nu_\mathrm{WNM}$ is the peak frequency of the template (i.e. 28.1\,GHz in flux density), $\nu_\mathrm{r}$ is the reference frequency to fit the flux density to, and $\nu_0$ is the fitted peak frequency. In this case we fit the flux density of the spinning dust curve at 28.4\,GHz. We also tried other models such as the log-normal approximation~\citep{stevenson2014paper} or multi-component \textsc{SpDust} models~\citep{planck_ame} but these models require more free parameters that are not favoured by the data.

We model the contribution of the cosmic microwave background (CMB) anisotropies using 
\begin{equation}
    S_\mathrm{cmb} = \frac{2 k \nu^2}{c^2} \Omega \Delta T , 
\end{equation}
where $\Delta T$ is the mean CMB brightness over the aperture in thermodynamic units\footnote{To convert Rayleigh-Jeans units ($T_\mathrm{RJ}$) to thermodynamic units ($T_\mathrm{K}$) we $T_\mathrm{RJ} = T_\mathrm{K} x^2 e^{x}/(e^{x} -1)^2$ where $x = h \nu / k T_\mathrm{cmb}$.}, and the other symbols have their standard definitions. 

\subsection{MCMC Fitting}
\label{sec:mcmc}

We use the \textsc{emcee}~\citep{emcee2019} implementation of the Markov Chain Monte Carlo (MCMC) to sample from our spectral model. Using an MCMC sampler to perform the model fitting is preferable to a more traditional least-squares method as it allows for a more complete interpretation of parameter correlations, and allows for the implementation of priors. We use 200 chains, each with 10000 samples, with a burn-in of 4000 samples for each spectral fit; these MCMC parameters were chosen as a balance between having a chain length that was sufficient for most chains to pass Gelman-Rubin convergence tests~\citep{Gelman1992} without using excessive computational resources. Any chains that had not converged after 10000 samples were reinitialised with starting parameters sampled from the posterior distribution of the converged chains and then rerun for 10000 samples. Finally, we tested for sample-sample autocorrelation and found that a thinning the chains by 15 samples was sufficient to remove sample-sample correlations.

Starting parameters for the MCMC chains are calculated using the \textsc{Python} \textsc{numpy} Levenberg-Marquardt least-squares implementation. We enforced positivity for all amplitude parameters (except for the CMB anisotropy amplitude). The impact of this prior is seen most in the determination of the emission measure (EM), which has a cut-off in its lower tail causing a slight underestimate of its uncertainty. The effect of the hard prior on EM can be seen in \autoref{fig:parameter_corr}.

We used the \textsc{Fastcc} package\footnote{\url{www.github.com/mpeel/fastcc}} \citep{peel2022_fastcc} to calculate colour corrections at the reference frequency given in \autoref{tab:fluxes} for each survey. The \textsc{Fastcc} code offers precalculated fits for a range of source spectral indices, utilising the measured bandpasses of several surveys, including: WMAP, \textit{Planck}, QUIJOTE, and \mbox{C-BASS}. We integrated the \textsc{Fastcc} code into our spectral fitting, updating the colour corrections for each survey for the trial spectrum derived by each sample in the MCMC chains. Colour corrections typically alter the observed flux densities 1--2\,per\,cent.

\subsection{The CMB}
\label{sec:cmb}

At microwave frequencies the brightness of M\,31 is comparable to the brightness of the CMB anisotropies at scales of a approximately $1^\circ$. In fact, between 70--100\,GHz approximately half of the total flux density within the aperture is related to a CMB fluctuation. Therefore even small errors in the contribution of the CMB fluctuations could result in a relatively large error in the flux density at microwave frequencies. As such, the CMB contribution to M\,31 must be carefully considered.

There are two ways to approach the CMB contribution: model the CMB anisotropies as an additional component in the spectral fitting, as was done in the \textit{Planck} analysis \citep{planck_andromeda}; or the CMB contribution can be subtracted from the maps using one of the \textit{Planck} CMB solutions. The first solution is problematic since it is adding an additional parameter to the overall fit, and the CMB contribution is also highly degenerate with the other components (synchrotron, AME, and free-free emission) at microwave frequencies.  The second solution of subtracting the CMB directly from the maps faces the question of the reliability of the various CMB solutions. When we include the CMB (best-fit parameters given in \autoref{tab:parameters}) we find we cannot easily constrain the CMB brightness within the aperture. We find the CMB brightness to be $0.3 \pm 1.1$\,$\mu$K in thermodynamic units, which is equivalent to $0.03\pm 0.12$\,Jy at 30\,GHz and $0.2\pm 1.0$\,Jy at 100\,GHz. We are not able to  constrain the CMB anisotropies directly from the integrated spectrum, and even the uncertainties from the differences in the four CMB models, discussed in \autoref{sec:uncertainties}, are larger than the fitted CMB brightness. Measuring the flux density of the CMB anisotropies within the M\,31 aperture using the CMB models gives a flux density of $0.2$--$0.3$\,Jy at 30\,GHz---an order of magnitude larger than the fit predicts. Further, the CMB amplitude is highly degenerate with the free-free and AME amplitudes, and it appears the flux density of the CMB anisotropies are being absorbed into the free-free emission measure parameter. It is for these reasons that we do not favour fitting for the CMB spectrally, but opt instead to subtract it using one of the available CMB models. 

\begin{figure}
    \centering
    \includegraphics[width=0.49\textwidth]{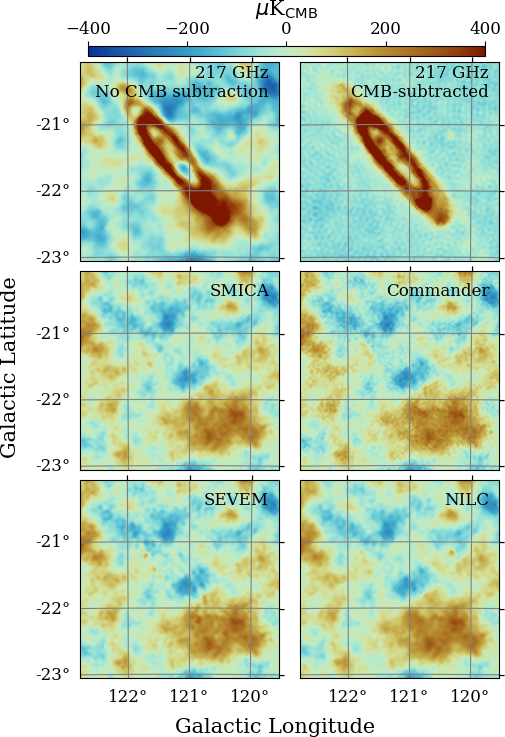}
        \caption{\textit{Top row}: \textit{Planck} 217\,GHz data before and after subtraction of the SMICA CMB solution. \textit{Middle row}: SMICA and Commander CMB solutions. \textit{Bottom row}: SEVEM and NILC CMB solutions. These maps are shown at 5\,arcmin resolution to show the M\,31 residuals in the CMB maps, however the fluctuations will be substantially smaller when smoothed to the $1^\circ$ resolution used in the analysis.}
    \label{fig:cmb}
\end{figure}

In \autoref{fig:cmb} we show a side-by-side comparison of the \textit{Planck} 217\,GHz map with the four \textit{Planck} CMB solutions: Independent Component Analysis of power spectra \citep[SMICA;][]{Cardoso2008}; needlet-based internal linear combination \citep[NILC;][]{Delabrouille2009}; Commander--a pixel-based parameter fitting method \citep{Eriksen2008}; and a multi-resolution internal template cleaning method \citep[SEVEM;][]{FernandezCobos2012}. There is a CMB fluctuation around the lower-right edge of the disk that is as bright as, and at many frequencies brighter than, the emission from M\,31 itself. It has been suggested it may not be a CMB fluctuation but may be associated with cold gas clouds in the M\,31 halo \citep{DePaolis2014,Tahir2022} though we did not consider this possibility for this analysis. The four different CMB solutions are very similar, although there are low-level residuals of the molecular disk of M\,31 in the SEVEM and NILC solutions. The SMICA and Commander solutions are very similar, with a negligible M\,31 residual. 

Comparing the aperture photometry of M\,31 when using each of the four different CMB maps shows that at 22.8\,GHz the largest change in flux density between all of the models is $\Delta S = 0.1$\,Jy, which is approximately 5\,per\,cent of the total flux density measured. At 70.4\,GHz, where the CMB is the dominant component, the largest difference is $\Delta S = 0.8$\,Jy between the CMB components---approximately 50\,per\,cent of the total measured flux density. We find that the Commander solution has the highest flux density, while the SMICA solution gives the lowest flux density. We therefore choose to use the SMICA solution to subtract the CMB from the WMAP and \textit{Planck} data. 

\subsection{Aperture Photometry}
\label{sec:photometrydetails} 

\autoref{tab:fluxes} presents the flux density and uncertainty, excluding colour corrections, at each frequency. We do not include colour corrections to enable direct comparison between our analysis and the \textit{Planck} \citep{planck_andromeda} and SRT \citep{battistelli_andromeda} analyses. The \textit{Planck} flux densities have not had the CMB or background sources subtracted, while the SRT flux densities have had both the CMB and background sources subtracted.  

We used aperture photometry to construct an integrated flux density spectrum of M\,31 using apertures matched to those used in the \textit{Planck} analysis \citep{planck_andromeda} and the more recent analysis by the SRT \citep{battistelli_andromeda}. We used an elliptical aperture encompassing the entirety of M\,31. The semi-major axis of the aperture is 100\,arcmin, with a minor-to-major ratio is 0.7, and a position angle of $45^\circ$ relative to Galactic North\footnote{The solid angle of the aperture was $\Omega = 3.72 \times 10^{-3}$\,sr}.

\begin{figure*}
 \centering
 \includegraphics[width=0.33\textwidth]{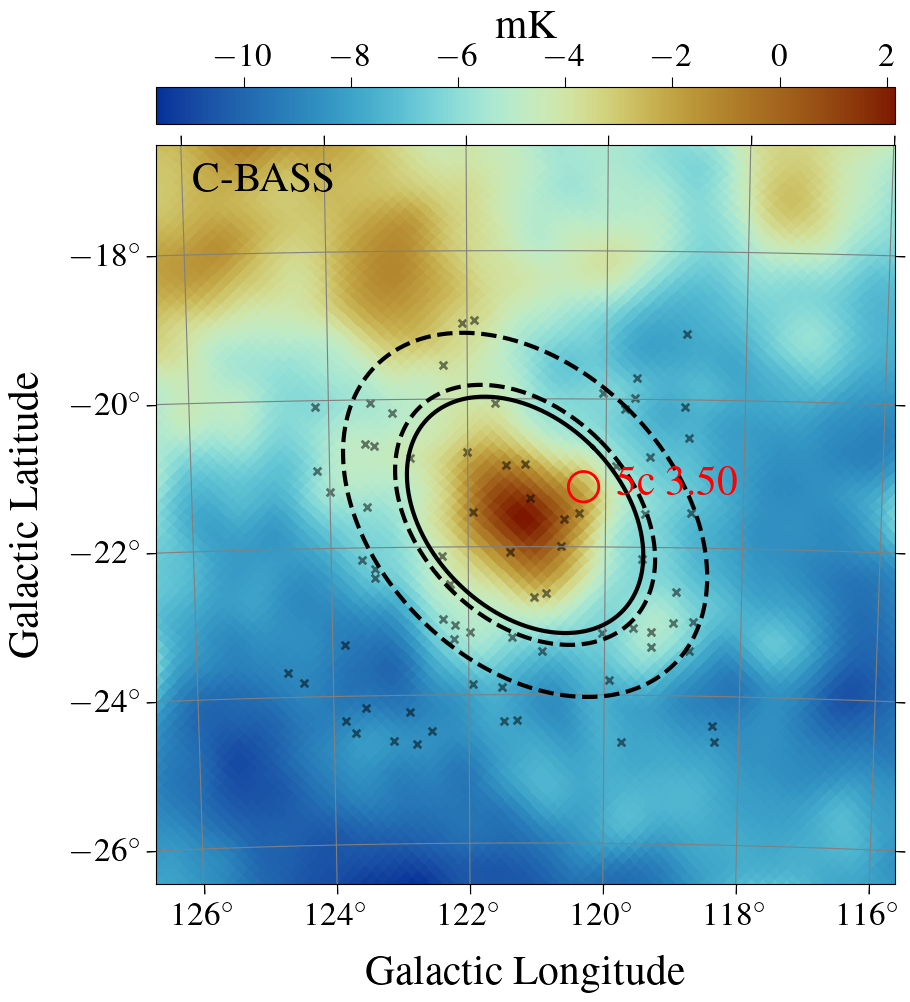}
 \includegraphics[width=0.33\textwidth]{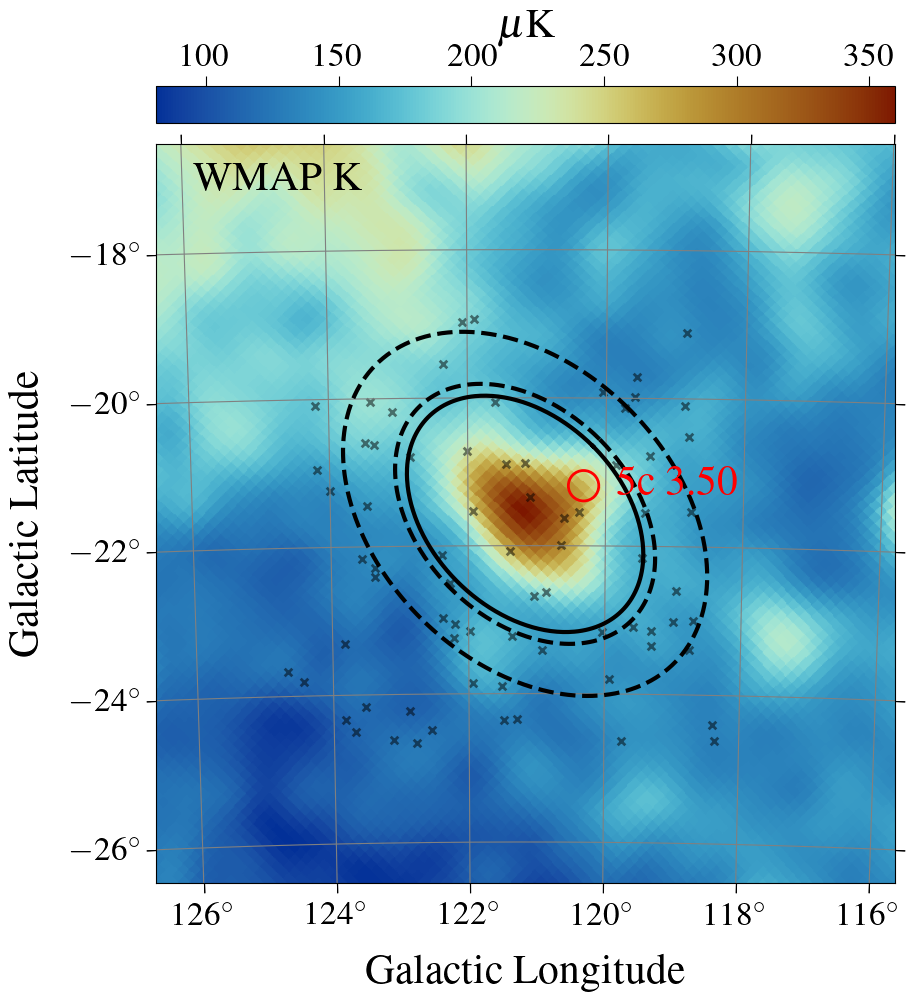}
 \includegraphics[width=0.33\textwidth]{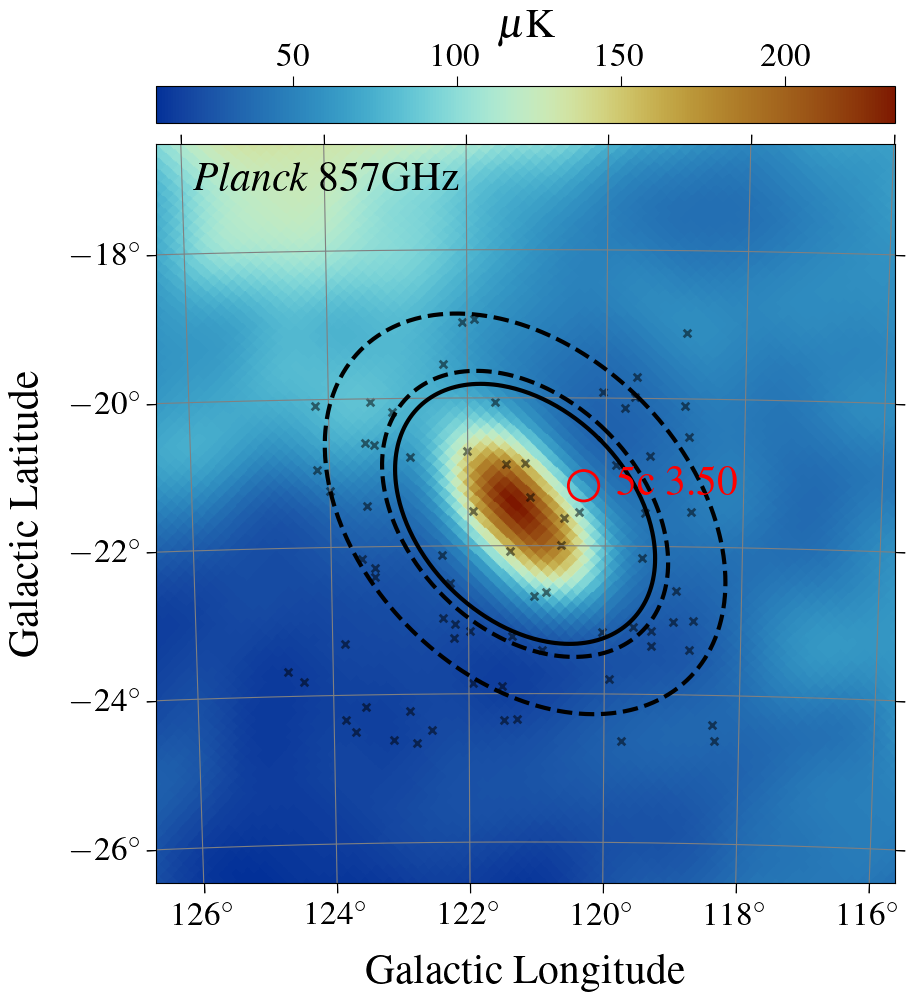}
 \caption{M\,31 as seen at 4.76\,GHz by \mbox{C-BASS} (\textit{left}), WMAP \textit{K-band} (22.8\,GHz, \textit{middle}), and \textit{Planck} 857\,GHz (\textit{right}), after smoothing to 1$^\circ$. The flux density was measured from pixels within the inner ellipse, while the background was estimated from the pixels between the inner and outer ellipses. The \textit{gray} markers indicate the location of known radio sources. The brightest source at 4.76 and 22.8\,GHz is 5C\,3.50, which we mark with a \textit{red} circle. All maps have had the SMICA CMB solution subtracted.}
 \label{fig:andromeda_apertures}
 \end{figure*}

To estimate the background emission, we determined the median of the pixels within an elliptical annulus with an inner and outer semi-major axis of 110\,arcmin and 154\,arcmin respectively. This annulus is 4\,arcmin wider than that used by the \textit{Planck} and SRT analyses. We left a 10\,arcmin gap between the aperture and annulus to minimize the over-subtraction of source flux density that was smoothed outside of the aperture. \autoref{fig:andromeda_apertures} shows the aperture and annulus used for three bands: \mbox{C-BASS}, WMAP \textit{K-band} (22.8\,GHz), and \textit{Planck} 857\,GHz.

\subsection{Photometry Uncertainties}
\label{sec:uncertainties}

The uncertainty, $\sigma_S^2$, in the measured flux density, $S$, has several contributions: the pixel noise uncertainty within the aperture, the uncertainty in the mean of the background annulus, the calibration uncertainty of the survey (given in \autoref{tab:fluxes}), the uncertainty in the background radio source brightnesses, and the background emission uncertainty. We define the uncertainty in the flux density at each frequency as
\begin{equation}\label{eqn:flux_uncertainties}
  \sigma_S^2 = \sigma_\mathrm{annu}^2 N_{S} \left[ 1 + \frac{\pi}{2}\frac{N_S}{N_\mathrm{annu}}\right] + S^2 \delta_g^2  + \sigma^2_\mathrm{src} + \sigma_\mathrm{cmb}^2 N_{S},
\end{equation}
 where $N_S$ and $N_\mathrm{annu}$ are the number of pixels in the aperture and annulus respectively, $\sigma_\mathrm{annu}$ is the standard deviation of the annulus, $\delta_g$ is the calibration uncertainty, $\sigma_\mathrm{src}$ is the uncertainty from the background radio sources, and $\sigma_\mathrm{cmb}$ is the uncertainty due to residuals after subtracting a model of the CMB.

 To estimate $\sigma_\mathrm{src}^2$ we used the model uncertainties associated with each source fit (as described in \autoref{sec:source-models}). We calculate the model variance for each source at each frequency and sum over all of these to create a source variance map. The source aperture variance ($\sigma_\mathrm{src}^2$) is simply the quadrature sum of the source variances within the aperture. 

There is no simple way to determine the correlated systematics errors associated with the four CMB solutions in the region around M\,31 (i.e., it is possible all four maps are subtracting some flux density associated with M\,31). However, we can use the differences between all four CMB solutions to obtain a lower limit on the CMB uncertainty, which we refer to as $\sigma_\mathrm{cmb}$ in \autoref{eqn:flux_uncertainties}. To do this we created the six possible unique difference maps between the SMICA, NILC, SEVEM, and Commander CMB solutions. We then considered two methods to estimate the uncertainty in the CMB solutions. First, we measured the integrated flux density of each difference map, and took the standard deviation between them. At 30\,GHz the typical uncertainty from the CMB was 0.041\,Jy, while at 70\,GHz it was 0.20\,Jy. In thermodynamic units the uncertainty in the CMB was 1.34\,$\mu$K. The second method was to measure the mean standard deviation of the pixels within the aperture for each difference pair. In this case we found the uncertainty in the CMB, near M\,31, was 1.55\,$\mu$K. For $\sigma_\mathrm{cmb}$, we used the average of these two uncertainties to get $\sigma_\mathrm{cmb}=1.45$\,$\mu$K, which is scaled to each frequency. As stated earlier, this will be an underestimate of the total CMB uncertainty as this method cannot account for correlated errors between all four CMB solutions.

In \autoref{tab:fluxes}, we show the measured flux densities with and without subtracting a CMB model. After subtracting the CMB, we find that the uncertainty in the flux density decreases. This is because the CMB fluctuations contributing to the uncertainty from within the annulus are greater than the additional uncertainty resulting from the differences in the CMB models. For instance, we note that in the frequency range of 70--100\,GHz, when the CMB is not subtracted, the uncertainties are approximately 3\,times larger than when the CMB is subtracted. This is because the CMB at these frequencies is the dominant source of emission. For example, at 30\,GHz the typical RMS due to the CMB anisotropies within the annulus is 0.11\,Jy, while the uncertainty due to CMB model subtraction is just 0.041\,Jy.

\section{Extragalactic Point Sources}
\label{sec:sources}

\subsection{Source Fitting and Subtraction}
\label{sec:source-models}

The spectral energy density (SED) of each source was modelled using a simple power-law, or a curved power-law model. Fits were performed on all available radio data between 0.1 and 100\,GHz. More details about each source and the source catalogue are given in \mbox{Appendix~\ref{sec:sources_catalogues}}. For sources modelled by a simple power-law relationship we used
\begin{equation}\label{eqn:powerlaw}
    f(\nu)=A\left(\frac{\nu}{\nu_r}\right)^{\alpha},
\end{equation}
where $A$ is the amplitude, $\nu_r$ is a fixed reference frequency at 1\,GHz, and $\alpha$ is the source spectral index. For sources with curved spectra we included a curvature term into the power-law fit, changing \autoref{eqn:powerlaw} to
\begin{equation}\label{eqn:curved_powerlaw}
    f(\nu)=A\left(\frac{\nu}{\nu_r}\right)^{\alpha + C \log(\nu)},
\end{equation}
where $C$ is the curvature term. The models were fit using the \textsc{SciPy}\footnote{\url{https://scipy.org/}} implementation of the L-BFGS-B algorithm \citep{Byrd1995} in the \textsc{minimize} package.

We used the Akaike information criterion \citep[AIC,][]{Akaike1974} to determine which model was preferred for each source. As we have a small number of measurements per source, we use the corrected AIC statistic that modifies the penalty function for small sample sizes (AICc). The AICc is defined as 
\begin{equation}
    \mathrm{AICc} = -2 \mathcal{L} + \frac{2 k^2 + 2 k}{n - k - 1}, 
\end{equation}\label{eqn:AIC}
where $\mathcal{L}$ is the maximum of the log-likelihood, $k$ is the number of model parameters, and $n$ is the number of measurements. We find that for the majority of sources a simple power law is sufficient, with several sources modelled using a curved power-law (see \autoref{tab:sources}). We used the model with the lowest AIC value as the best-fitting model for each source.

For many sources in the source catalogue (Appendix~\ref{sec:sources_catalogues}) the uncertainty in the catalogue was either very small (less than 1\,per\,cent) or was not given at all. In these instances we set the uncertainty to be 10\,per\,cent of the flux density. 

\subsection{5C3.50}
\label{sec:5c0350}

\begin{figure}
    \centering
    \includegraphics[width=0.49\textwidth]{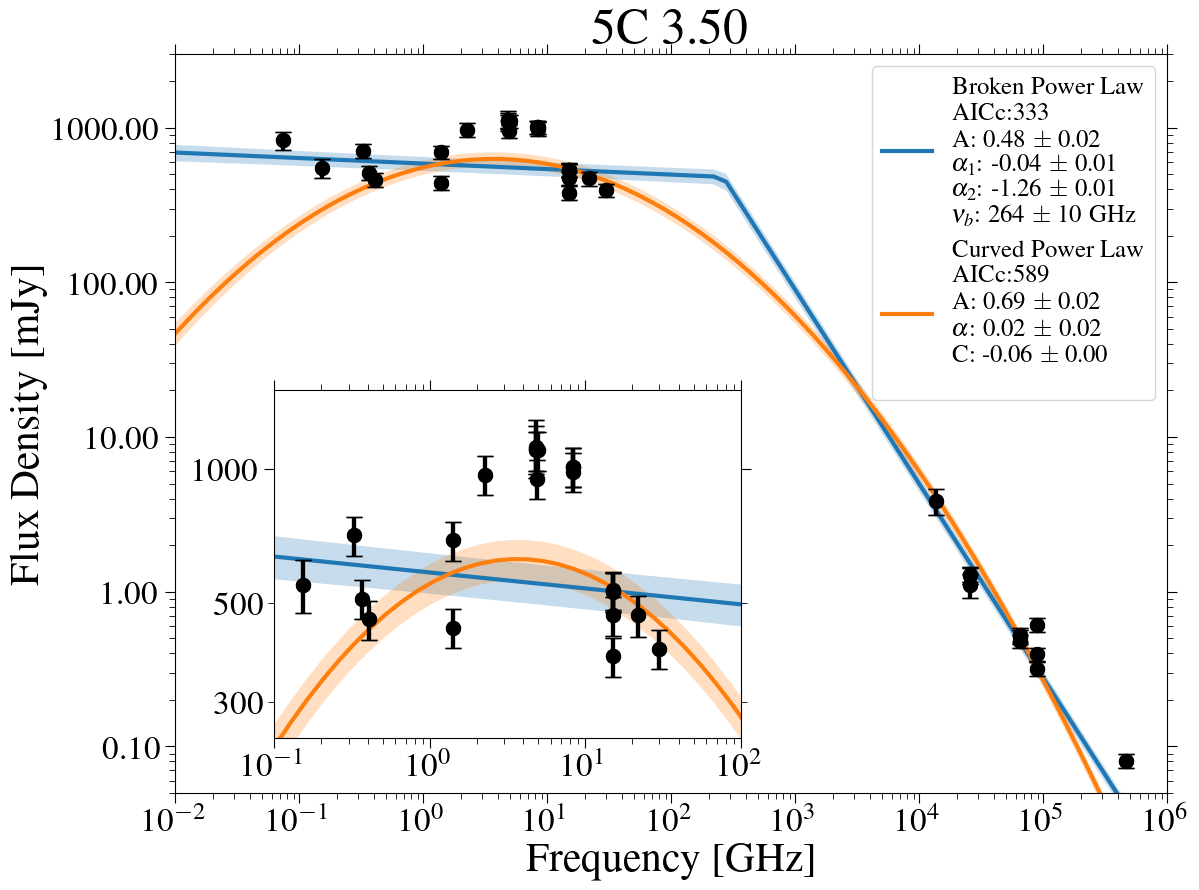}
    \caption{Ancillary data on 5C\,3.50 (\mbox{B3\,0035+413}), the brightest extragalactic background source near M\,31. Fits for the broken power-law (\textit{blue}) and curved power-law (\textit{orange}) models are shown with the shaded areas indicating the 1$\sigma$ model uncertainties. The inset shows a zoom-in of the radio data between 0.1--100\,GHz. The legend shows the AICc and best-fit parameters for the curved power-law (\autoref{eqn:curved_powerlaw}) and broken power-law (\autoref{eqn:broken_powerlaw}) models.}
    \label{fig:point_sources}
\end{figure}

The brightest extragalactic background source at frequencies \mbox{5--70\,GHz} is the AGN 5C\,3.50~\citep[referred to as \mbox{B3\,0035+413} in][]{battistelli_andromeda}. In \autoref{fig:point_sources} we show the spectrum of \mbox{5C\,3.50} from radio to infrared frequencies. We can see that at frequencies less than  100\,GHz the source has a flat spectrum and is steeply falling at higher frequencies. The peak flux density of the source is $\approx 0.6$\,Jy at 5\,GHz, which is about 15\,per\,cent of the total flux density within the aperture at 5\,GHz, and similarly it accounts for 24\,per\,cent of the total flux density at 30\,GHz. Therefore it is critical to determine an accurate model of the source flux density.

Complicating the model of \mbox{5C\,3.50} are indications of variability, which is to be expected from a compact AGN. Evidence of the variability of \mbox{5C\,3.50} can be seen in the inset spectrum of \autoref{fig:point_sources}, where we can see between 4 and 8\,GHz there is a jump in flux density by a factor of two. We also see evidence for variability within frequency bands, for example the flux density for \mbox{5C\,3.50} at 1.4\,GHz measured by the Very Large Array \citep[VLA,][]{condon1998} was $0.692 \pm 0.021$\,Jy in 1998, while the flux density reported by the Green Bank Telescope \citep[GBT,][]{White1992} was $0.404$\,Jy (no uncertainty was reported) in 1992; a 50\,per\,cent change in a 6 year period.

\begin{figure}
    \centering
   \includegraphics[width=0.49\textwidth]{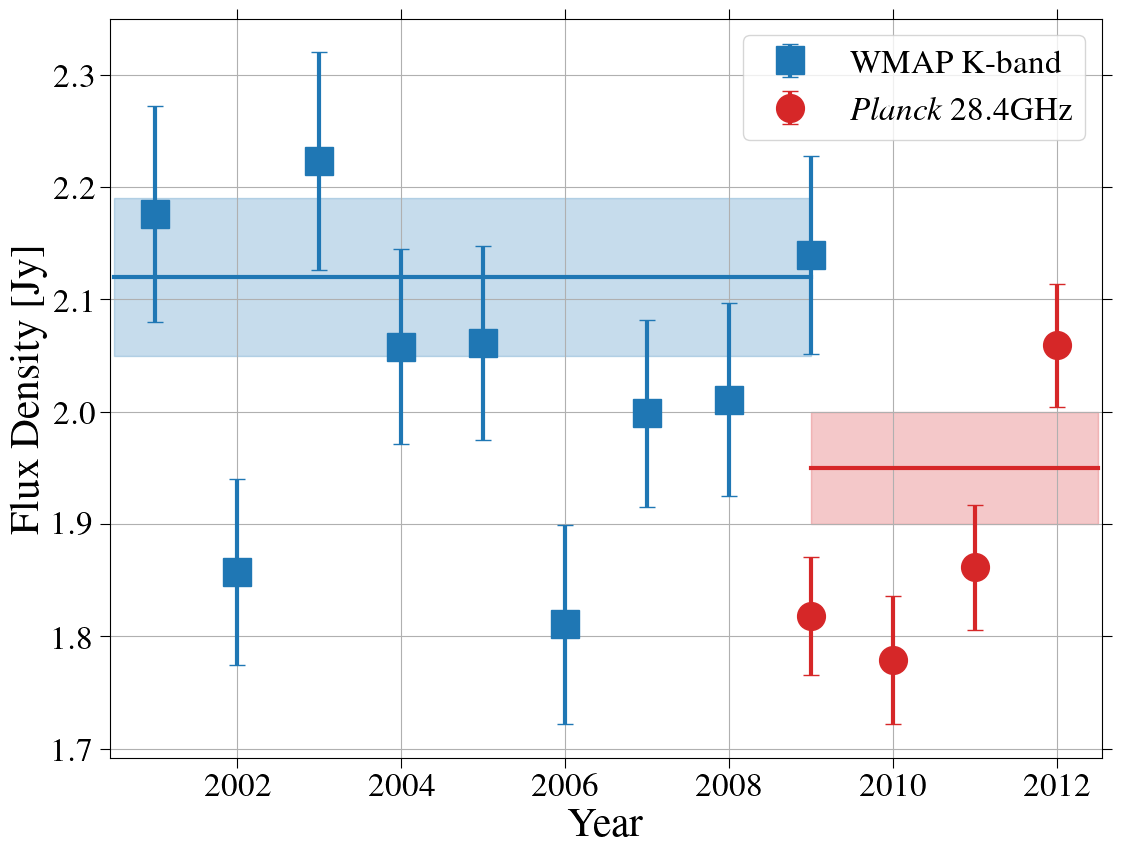}
    \caption{Integrated flux density of M\,31 (including \mbox{5C\,3.50}) from WMAP 22.8\,GHz and \textit{Planck} 28.4\,GHz data for each year between 2001 and 2012. We use the same aperture photometry parameters as with the main analysis (described in \autoref{sec:photometry}) but do not subtract nearby sources. The variations in the flux density is a combination of measurement error and true variations in the flux density of \mbox{5C\,3.50}. The solid lines and highlighted regions show the measured flux densities at 22.8 and 28.4\,GHz from \autoref{tab:fluxes}.}
    \label{fig:wmap_planck_yearly}
\end{figure}

To assess the variability of \mbox{5C\,3.50}, we used historical data covering 2009--2019 from the OVRO 40-m monitoring program \citep{OVRO_2011}. The data shows a steady decline in brightness from 0.59$\pm$0.01\,Jy in 2009 to a minimum of 0.32$\pm$0.01\,Jy in 2019. This decline covers the period when most observations used in this analysis were in progress, but does not include the period when WMAP data were being taken (2001--2010). To observe the effect this source has on the data, we looked at the single-year data from WMAP and found that, on average, most years are within 1$\,\sigma$ of the survey average flux density, but there were significant drops in flux density in 2002 and 2006. The change in flux density in 2002 and 2006 were $\Delta S_{2002} = -0.26 \pm 0.11$\,Jy and $\Delta S_{2006} = 0.30 \pm 0.11$\,Jy, respectively. If these drops in the flux density of M\,31 were associated with \mbox{5C 3.50} then that would imply a change in the flux density of \mbox{5C 3.50} of 40\,per\,cent relative the model average given in \autoref{tab:sources}. The yearly data from the \textit{Planck} 28.4\,GHz data (2009 to 2012) also shows that there is significant evidence for variations in the brightness of \mbox{5C\,3.50}, its flux density changing slowly between 2009 and 2012 by 30--50\,per\,cent---similar to the largest changes seen within the WMAP data. \autoref{fig:wmap_planck_yearly} shows the measured flux densities from WMAP and \textit{Planck} for each year. Overall, the impact of the variability of \mbox{5C\,3.50} over the most critical frequency range of \mbox{10--100\,GHz} is significant but does not exceed 10\,per\,cent of the total flux density of M\,31. At 30\,GHz this would constitute a maximum flux density change of approximately 0.2\,Jy. 

For \mbox{5C\,3.50} we fit the SED over all available radio and infrared data. We do not fit over the narrower range of frequencies because of the broad peak at 4--8\,GHz that is shown in the inset panel of \autoref{fig:point_sources}, which may be from source variability over the ten year time span of all the radio observations. It is also possible that the peak is a real structure in the spectrum of \mbox{5C\,3.50} due to processes such as synchrotron self-absorption, but further observations would be needed to confirm this. We fit the SED of \mbox{5C\,3.50} for both the curved power-law model (\autoref{eqn:curved_powerlaw}) and an additional broken power-law model defined as:
\begin{equation}\label{eqn:broken_powerlaw}
    f(\nu)=A\left(\frac{\nu}{\nu_b}\right)^{\alpha(\nu)},
\end{equation}
where $\alpha(\nu)$ at frequencies less than the break frequency, $\nu_b$, are $\alpha_1$ and $\alpha_2$ at frequencies greater than $\nu_b$. The functional form of $\alpha(\nu)$ was 
\begin{equation}
    \alpha(\nu) = \alpha_1 + S(\nu-\nu_b) \left(\alpha_1 - \alpha_2\right),
\end{equation}
where $S$ is a sigmoid function defined as $S(x)=(1 + e^{-x})^{-1}$. Using this model we find that at radio frequencies below the best-fit break frequency ($\nu_b = 264 \pm 14$\,GHz) the spectrum can be described by a power-law with a spectral index of $\alpha_1 = -0.036 \pm 0.010$, and at higher frequencies the source has a steeper spectral index of $\alpha_1 = -1.25 \pm 0.010$. To account for the variability of \mbox{5C\,3.50} we add an additional 10\,per\,cent uncertainty at each frequency, which is the dominant source of uncertainty shown in the models in \autoref{fig:point_sources}.

\section{Results}
\label{sec:results}

\begin{figure*}
    \centering
    \includegraphics[width=0.85\textwidth]{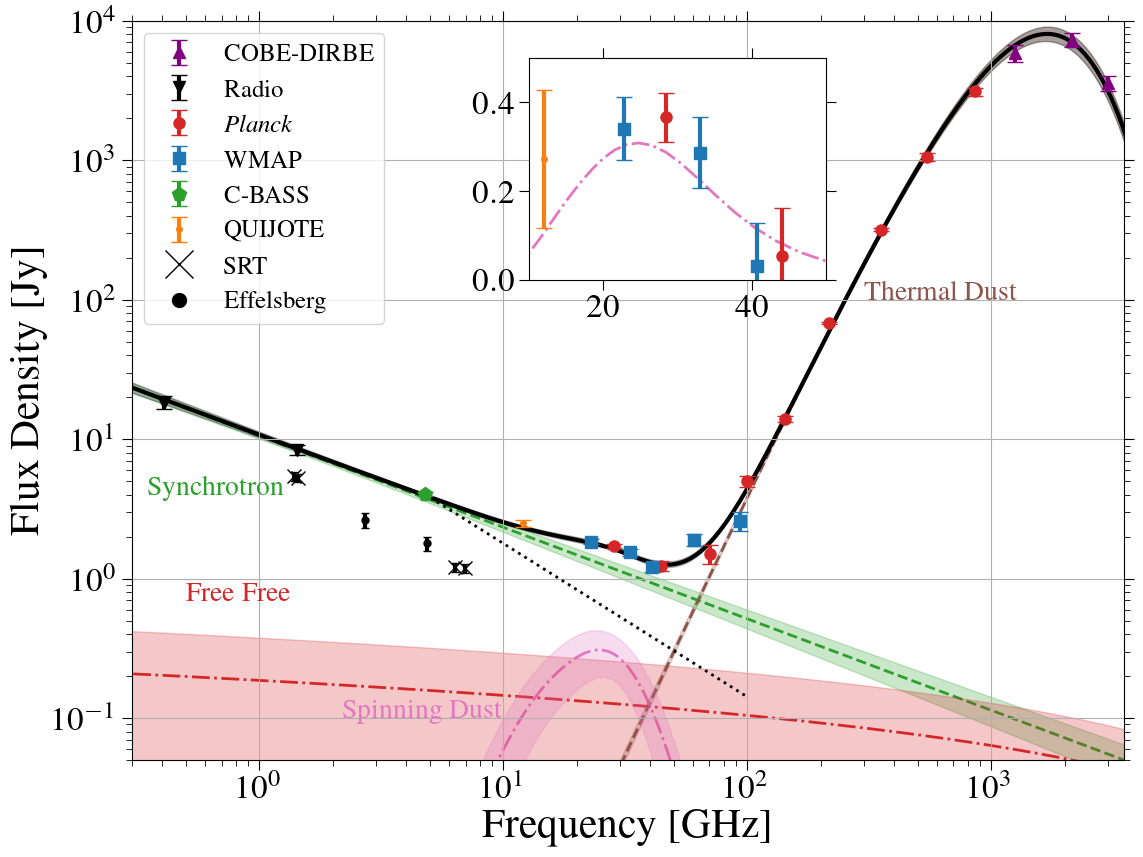} \\
   \caption{Integrated flux density spectrum for the M\,31 after subtracting the SMICA CMB solution and background radio sources. Four emission components are fitted for: synchrotron, free-free, thermal dust, and spinning dust emission. The inset spectrum shows the residual flux density after subtracting all but the spinning dust model in the region around the fitted spinning dust peak frequency. We include the flux densities measurements from the SRT \citep{battistelli_andromeda} and the Effelsberg \citep{Beck2020} observations of M\,31 for reference but these were not used in the fit. The \textit{black-dotted} line shows the best-fit synchrotron spectral index from \citet{battistelli_andromeda} with the amplitude referenced to the \mbox{C-BASS} flux density measurement at 4.76\,GHz.}
\label{fig:andromeda_sed}
\end{figure*}

We present the best-fitting spectral energy distribution (SED) after subtracting the SMICA CMB solution and background radio sources in \autoref{fig:andromeda_sed}. The best-fitting parameters are listed in \autoref{tab:parameters}, alongside those given in \citet{battistelli_andromeda} and \citet{planck_andromeda}. Our preferred model is the "AME \& power-law synchrotron" model in \autoref{tab:parameters}. For this section and \autoref{sec:discussion}, unless explicitly stated, all comparisons between models will be between our preferred model and the other models listed in \autoref{tab:parameters}. 

 In \autoref{fig:andromeda_sed} we can clearly see that at all frequencies spinning dust emission is sub-dominant to the synchrotron component. We find that when fitting for spinning dust emission we find a 3$\sigma$ detection of $A_\mathrm{AME,30} =~$\CMBSRCsubSpAAme\,Jy, which is a  similar to significance to that of the \textit{Planck} analysis \citep{planck_andromeda} but is much less significant than the SRT measurements \citep{battistelli_andromeda}. The biggest difference between this analysis and the SRT analysis is the constraint of the lower frequency synchrotron emission. We can see that the SRT measurements at 6.313\,GHz are four times fainter than the \mbox{C-BASS} measurements at 4.76\,GHz in \autoref{tab:fluxes}. The substantially steeper estimate of the M\,31 synchrotron spectrum from the SRT models therefore predicts a far larger spinning dust component to make up for the lower synchrotron emission at 30\,GHz. However, with \mbox{C-BASS} the predicted synchrotron spectrum is closer to that of previous studies, and the spinning dust is similarly small. We believe that the main reason for the differences between the SRT and \mbox{C-BASS} measurements is due to spatial filtering in the SRT data, this is discussed more in \autoref{sec:discussion}.

\begin{table*}
        \centering
            \caption{Mean posterior parameter estimates for the four cases discussed in the text. For comparison we include the previous parameter fits from the \textit{Planck} and SRT analyses. }
        \begin{tabular}{lcc|ccc|cc}
        \hline
                  &              &            & \multicolumn{3}{|c|}{CMB \& Source Subtracted} &  \\
                  & No CMB       & CMB        & AME \& Power-Law$^{1}$ & AME \& Curved & No AME \& Power-Law & & \\
        Parameter & Subtraction  & Subtracted & Synchrotron        & Synchrotron     & Synchrotron & \textit{Planck} 2015 & SRT 2019 \\
        \hline
        \hline

$A_\mathrm{sync}$ (Jy) & \noCMBsubASync & \CMBsubASync & \CMBSRCsubASync & \CMBSRCsubcurvedASync & \CMBSRCsubnoameASync & \planckmodelASync & \srtmodelASync\\
$\alpha$ & \noCMBsubAlpha & \CMBsubAlpha & \CMBSRCsubAlpha & \CMBSRCsubcurvedAlpha & \CMBSRCsubnoameAlpha & \planckmodelAlpha & \srtmodelAlpha\\
$C$ & \noCMBsubCsCurve & \CMBsubCsCurve & \CMBSRCsubCsCurve & \CMBSRCsubcurvedCsCurve & \CMBSRCsubnoameCsCurve & \planckmodelCsCurve & \srtmodelCsCurve\\
$\mathrm{EM}$ (cm$^{-6}$\,pc) & \noCMBsubEm & \CMBsubEm & \CMBSRCsubEm & \CMBSRCsubcurvedEm & \CMBSRCsubnoameEm & \planckmodelEm & \srtmodelEm\\
$\delta_{T}$ ($\mu$K) & \noCMBsubDt & \CMBsubDt & \CMBSRCsubDt & \CMBSRCsubcurvedDt & \CMBSRCsubnoameDt & \planckmodelDt & \srtmodelDt\\
$T_d$ (K) & \noCMBsubTD & \CMBsubTD & \CMBSRCsubTD & \CMBSRCsubcurvedTD & \CMBSRCsubnoameTD & \planckmodelTD & \srtmodelTD\\
$\tau$ & \noCMBsubTau & \CMBsubTau & \CMBSRCsubTau & \CMBSRCsubcurvedTau & \CMBSRCsubnoameTau & \planckmodelTau & \srtmodelTau\\
$\beta$ & \noCMBsubBeta & \CMBsubBeta & \CMBSRCsubBeta & \CMBSRCsubcurvedBeta & \CMBSRCsubnoameBeta & \planckmodelBeta & \srtmodelBeta\\
$A_\mathrm{AME,30}$ (Jy) & \noCMBsubSpAAme & \CMBsubSpAAme & \CMBSRCsubSpAAme & \CMBSRCsubcurvedSpAAme & \CMBSRCsubnoameSpAAme & \planckmodelSpAAme & \srtmodelSpAAme\\
$\nu_\mathrm{AME}$ (GHz) & \noCMBsubSpNuAme & \CMBsubSpNuAme & \CMBSRCsubSpNuAme & \CMBSRCsubcurvedSpNuAme & \CMBSRCsubnoameSpNuAme & \planckmodelSpNuAme & \srtmodelSpNuAme\\
$\chi^2$ & \noCMBsubChisq & \CMBsubChisq & \CMBSRCsubChisq & \CMBSRCsubcurvedChisq & \CMBSRCsubnoameChisq & \planckmodelChisq & \srtmodelChisq\\
$\chi^2_r$ & \noCMBsubChisqReduced & \CMBsubChisqReduced & \CMBSRCsubChisqReduced & \CMBSRCsubcurvedChisqReduced & \CMBSRCsubnoameChisqReduced & \planckmodelChisqReduced & \srtmodelChisqReduced\\
AIC & \noCMBsubAicc & \CMBsubAicc & \CMBSRCsubAicc & \CMBSRCsubcurvedAicc & \CMBSRCsubnoameAicc & \planckmodelAicc & \srtmodelAicc\\ \hline
    \end{tabular}
    \begin{tablenotes}
    \item $^1$ This is the preferred model that we refer to most often in the main text.
    \end{tablenotes}
\label{tab:parameters}
\end{table*}
 
\begin{figure*}
    \centering
    \includegraphics[width=0.85\textwidth]{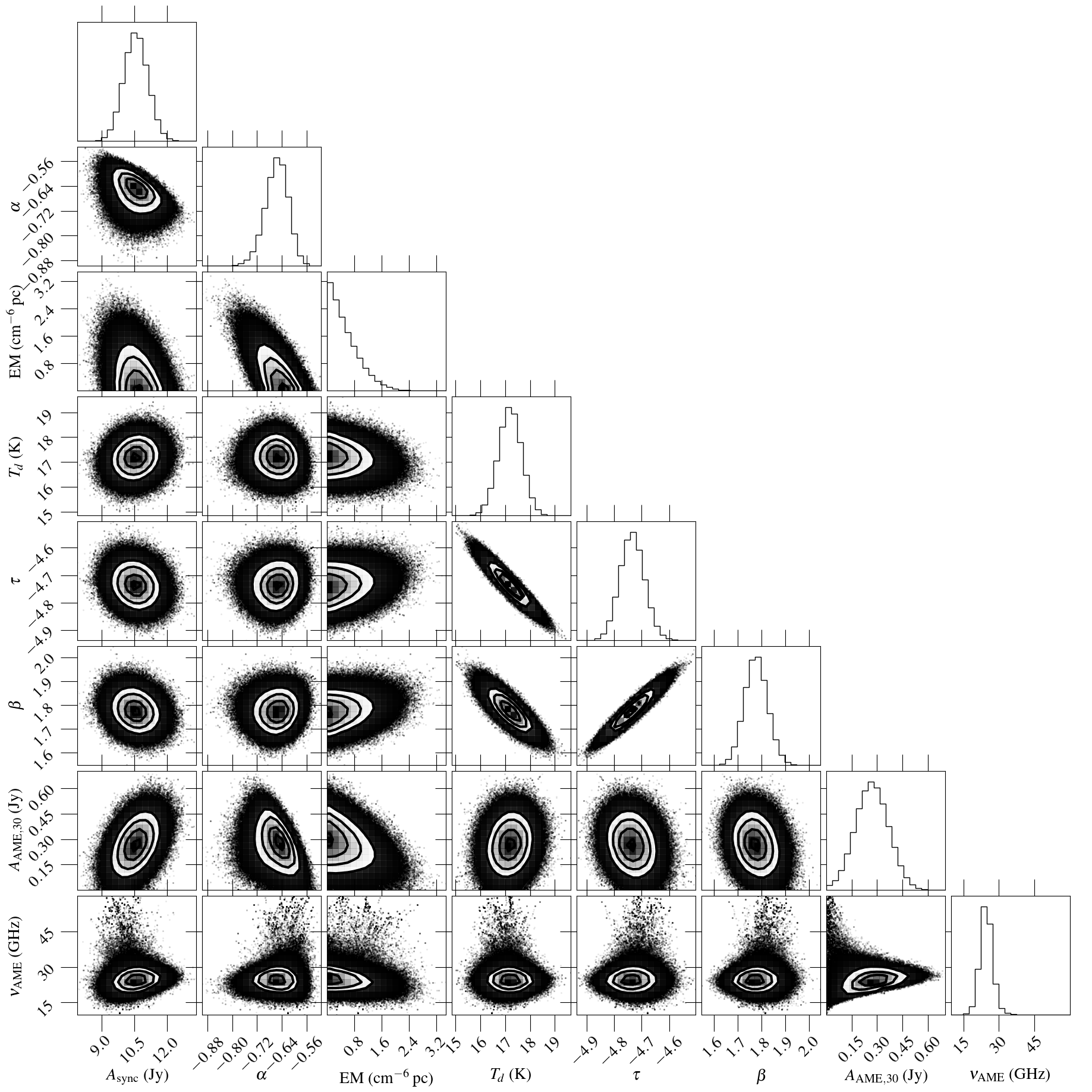}
   \caption{Posterior distributions for each parameter for the model shown in \autoref{fig:andromeda_sed}. The model includes an AME component and a power-law synchrotron component, and has both CMB and background sources subtracted. The hard cut-off in the EM parameter is due to a positivity prior.}
\label{fig:parameter_corr}
\end{figure*}

In \autoref{fig:parameter_corr} we show the posterior distributions of each parameter for the model with the CMB and sources subtracted, AME included, and a power-law synchrotron model. The figure shows that for most parameters the posterior is peaked. The emission measure (EM) parameter however shows a sharp cut-off due a hard positivity prior we enforce. As such, the EM uncertainties are underestimated in \autoref{tab:parameters} for all model configurations, but it is especially underestimated for the model shown in \autoref{fig:andromeda_sed} as we do not make a significant detection.

\autoref{fig:parameter_corr} also shows that there is a correlation between the EM, the synchrotron parameters ($A_\mathrm{sync}$ and $\alpha$) and the spinning dust amplitude ($A_\mathrm{AME,30}$). This is expected, and was also seen in \citet{planck_andromeda}, because a lower amplitude, steeper spectrum synchrotron component would require additional contributions from either the EM or spinning dust components to explain the flux density observed around $20$--$70$\,GHz. The strong correlation between the thermal dust parameters in  \autoref{fig:parameter_corr} is the well-known dust temperature--$\beta$ correlation caused by combination of effects such as noise uncertainties, multiple dust grain environments, and the physics of the dust grain emission \citep[e.g.,][]{Shetty2009,Ysard2012}.

It is possible to model AME using a log-normal distribution as has been shown in several analyses \citep{stevenson2014paper,Cepeda-Arroita2021}. However, fitting a log-normal distribution requires the introduction of an additional width parameter. We found that when using this model the width parameter was  highly degenerate with the EM parameter, implying that the data do not support this additional parameter.

We find that the spectral index of the synchrotron emission is flatter ($\alpha =~$\CMBSRCsubAlpha) than the spectral indices in \citet{battistelli_andromeda} or \citet{planck_andromeda}, and is more typical of other galaxies \citep{condon1998}. This is much flatter than the synchrotron spectrum found in our Galaxy, which typically has a spectral index of $\alpha = -1$ at frequencies of $\nu > 5$\,GHz \citep[e.g.,][]{Harper2022}. However, nearer the Galactic plane there is evidence that synchrotron emission is flatter \citep[e.g.][]{kogut2007,Fuskeland2014}, which will be the dominant component when measuring the integrated emission from a galaxy. 

In \autoref{tab:parameters} we compare the $\chi^2$ residuals and the Akaike Information Criterion \citep[AIC,][]{Akaike1974} of each model. As we are using nested models in the case of either including or excluding the spinning dust component (for columns under the "CMB \& Source Subtracted" heading in \autoref{tab:parameters}) the AIC is an ideal tool for comparing the models. Instead of simply using the maximum likelihood value to determine the AIC we sum over the entire posterior to better account for biases due to non-Gaussian posterior distributions \citep{Hilbe2017}. We define the modified AIC as before in \autoref{eqn:AIC}. Although the AIC does not give a direct indication of the absolute quality of the fit, values can be compared between two nested models, with the lower AIC value indicating the model is a better representation of the data.

The AIC values in \autoref{tab:parameters} show that fitting for AME after subtracting the CMB and point sources is slightly preferred ($\mathrm{AIC} =~$\CMBSRCsubAicc) over not including a spinning dust component ($\mathrm{AIC} =~$\CMBSRCsubnoameAicc). Taking the difference of the AIC values for with and without spinning dust gives $\Delta \mathrm{AIC} = 13$. In general, a  $\Delta \mathrm{AIC}  > 4$ is considered significant \citep{Hilbe2017}, implying that we do detect spinning dust in the integrated spectrum of M\,31. 

We also tried fitting for spectral curvature in the spectral index of the synchrotron emission from M\,31 using the same model given by \autoref{eqn:curved_powerlaw}. We found that the curved synchrotron model had the same AIC and $\chi^2$ values as the non-curved power-law spectrum. Fitting for a curved synchrotron component does not significantly change the AME component of M\,31, but does double the observed emission measure. Given theoretical insights, particularly for our own Galaxy, that there is little evidence for synchrotron steepening or flattening with frequency \citep[e.g.][]{Strong2011} and that the curvature term was detected with low significance ($C=~$\CMBSRCsubcurvedCsCurve), we prefer the model that includes no steepening. We give the best-fitting parameters including for the curved synchrotron component in \autoref{tab:parameters} for reference. 

\section{Discussion}
\label{sec:discussion}

\subsection{Comparison of C-BASS with Other Radio Observations}\label{sec:comparison_data}

The results presented in \autoref{sec:results} are driven by the introduction of the C-BASS point at 4.76\,GHz, where we measure a flux density of \cbassFluxDensity\,Jy. This is higher than the flux densities measured by the Effelsberg telescope \citep{berkhuijsen2003} at 4.85\,GHz of 1.79$\pm$0.20\,Jy and the SRT \citep{battistelli_andromeda} at 6.313\,GHz of $1.207\pm0.084$\,Jy, as shown in \autoref{fig:andromeda_sed}.

The most likely reason for the difference in the flux densities between \mbox{C-BASS} and the other two surveys is the spatial filtering over a small area of the sky. The SRT and Effelsberg observations have  resolutions of a few arcminutes, while \mbox{C-BASS} has 45\,arcmin resolution. In general, it is difficult to measure the largest scales observed by a survey due to large-scale systematics from the receiver, atmosphere or surrounding environment. In this case, the largest scales measured by the SRT and Effelsberg are comparable to the size of M\,31, as such these scales are more likely to be lost during data processing. Meanwhile, \mbox{C-BASS} observes the entire Northern hemisphere, and so preserving signal on scales of just a few degrees is considerably easier. We have verified this using end-to-end simulations to assess the impact of the \mbox{C-BASS} data processing and map-making on filtering the signal in the \mbox{C-BASS} map and found that there is no significant filtering beyond the first few multipoles and certainly none at scales of M\,31 (the details of these simulations will be given in the upcoming \mbox{C-BASS} survey paper). Therefore, the difference in flux density is either because both the SRT and Effelsberg are filtering out emission from M\,31, or possibly that \mbox{C-BASS} is adding in emission from our Galaxy that is filtered out in the higher resolution datasets. The latter is unlikely to be the case as we do not observe any bright Galactic background features in any of the lower resolution datasets and we effectively filter out scales larger than M\,31 by subtracting a tight background annulus, as shown by the \mbox{C-BASS} data in \autoref{fig:andromeda_apertures}. However, evidence for filtering can be seen in the SRT data around bright point sources such as \mbox{5C\,3.50} where a large negative ring surrounds the source.

To make a fair comparison with the lower resolution surveys requires replicating the filtering applied to the higher resolution data, which was attempted by \citet{battistelli_andromeda}. This process can only be approximated, as the higher resolution information is not available in the low resolution surveys. By using \mbox{C-BASS} data, with a comparable resolution to these other surveys, we can make a fairer comparison of the total integrated flux density of M\,31 without the need to make additional corrections.

\subsection{Synchrotron Emission}\label{sec:synchrotron}

We find that, for our favoured model including an AME component, a power-law synchrotron component, and with the CMB and background sources subtracted, the average synchrotron spectral index of M\,31 is \mbox{$\alpha =~$\CMBSRCsubAlpha}. Synchrotron emission is the dominant emission component around 30\,GHz---approximately twice the thermal free-free emission and AME contributions. The \textit{Planck} analysis in \citet{planck_andromeda} found a slightly steeper spectral index of $\alpha = -0.9 \pm 0.2$, however the inclusion of the \mbox{C-BASS} and QUIJOTE measurements show that the synchrotron spectrum is slightly flatter.

The 5\,arcmin resolution analysis of M\,31 using Effelsberg data at 2.645, 4.86, and 8.35\,GHz \citep{Beck2020} found a spectral index of $\alpha = -0.81 \pm0.02$ for the non-thermal emission component---a value that lies between our analysis and that of the \textit{Planck} analysis. In \autoref{fig:andromeda_sed}, we can see that the flux densities for M\,31 observed by the Effelsberg data are approximately a factor of two lower than those predicted by the best-fit model using the \mbox{C-BASS}/QUIJOTE data. Comparing the synchrotron spectral indices from the Effelsberg and our analyses suggests that the estimate of the synchrotron spectral index from the 64-m SRT observations of $\alpha = -1.1 \pm 0.09$ \citep{battistelli_andromeda,Fatigoni2021} is too steep---though all three are consistent with the \textit{Planck} result of $\alpha = -0.92 \pm 0.16$ within $2\sigma$. This is critical because a steeper synchrotron spectral index would imply a much lower synchrotron contribution at 30\,GHz that in the SRT model is replaced by the large AME contribution predicted by the SRT model.

In the Milky Way, estimates of the spectral index of diffuse synchrotron emission are generally found in the range \mbox{$-1.1 < \alpha < -1$} at WMAP and \textit{Planck} frequencies \citep[e.g.,][]{Davies2006,Dunkley2009,Harper2022}, and flatten to $\beta \approx -2.7$ at frequencies of a few hundred MHz to 1\,GHz \citep[e.g.][]{Lawson1987,reich1988}. At higher frequencies it is possible that further steepening of the synchrotron spectrum can occur due to spectral aging of the cosmic ray electron (CRE) population \citep{Strong2011}, though observing spectral aging in diffuse synchrotron emission is challenging; the effect of spectral aging has been clearly seen in supernova remnants \citep[e.g.,][]{Sun2011}. To test for spectral curvature in M\,31 we included a curvature term, shown in \autoref{eqn:powerlaw_curvature}. We found that when fitting the CMB and source subtracted M\,31 spectrum with a three component fit of: free-free, thermal dust, and curved synchrotron emission the curvature term is $C =~$\CMBSRCsubnoamecurvedCsCurve, implying there is no significant detection of curvature. If we include AME in the fit, then the curvature term is  $C =~$\CMBSRCsubcurvedCsCurve~reducing the synchrotron contribution near 30\,GHz; the excess flux density is then absorbed into the free-free component (but the AME amplitude remains unchanged). However, we find there is a $\approx 1\sigma$ constraint on the curvature, indicating that even an insignificant curvature in the M\,31 synchrotron spectrum can have a significant impact on the estimation of other parameters---especially the free-free contribution.

%
%
%
%
%
%
%
%
%
%
%

\subsection{Anomalous Microwave Emission in M31}

We find a $3$\,$\sigma$ detection of AME in M\,31 of \CMBSRCsubSpAAme\,Jy at 30\,GHz for the model where CMB and background sources are subtracted, and there is no synchrotron spectral curvature. Though the exact significance changes slightly between models, in all cases where the spinning dust model is included (and the CMB is subtracted) we detect AME at similar levels of significance. For the case where the CMB is not subtracted but instead fitted in the spectrum, we find that there is a barely significant detection ($1.2 \sigma$) of AME at \noCMBsubSpAAme\,Jy at 30\,GHz. This reduction in significance when not subtracting the CMB is because the frequency range where AME peaks is the same as the frequencies where the CMB is the dominant component, \mbox{30--100}\,GHz.

We find an AME amplitude that is less than the amplitude predicted from the \citet{planck_andromeda} analysis of $S_\mathrm{AME,30} = 0.7\pm0.3$\,Jy at 30\,GHz, and is much less than the predicted integrated AME flux density from the SRT analysis of $S_\mathrm{AME,25}=1.4\pm0.2$\,Jy at 25\,GHz. The inclusion of the 4.76\,GHz \mbox{C-BASS} data is the key difference between these analyses as it allows for the synchrotron emission at high frequencies to be constrained as discussed in \autoref{sec:synchrotron}.

We find that the AME brightness at 30\,GHz relative to the dust opacity at 353\,GHz, often referred to as AME emissivity, has a value of \emissivity--assuming a uniform brightness over our entire nominal aperture. This is much smaller than the AME emissivity found in Galactic sources with well-measured AME, which typically has a value in the range \mbox{5--10\,K/$\tau_{353}$} \citep{Hensley2016,Harper2022} for cirrus regions, but can be as high as $23.9$\,K/$\tau_{353}$ in particular molecular clouds \citep{planck_xx}.

We would expect that the average AME emissivity in M\,31 to be lower than what is found in targeted observations of AME sources. This is because we do not expect that AME to be uniformly distributed throughout the ISM in M\,31 or any galaxy. Moreover, the integrated spectrum of M\,31 is dominated by bright synchrotron and free-free features, such as supernova remnants or H\textsc{ii} regions, associated with the molecular disk. The integrated emission from H\textsc{ii} regions is rarely found to be dominated by AME \citep[e.g.,][]{Scaife2008,Paladini2015,Rennie2022}. Instead, we are more likely to find AME associated with dust clouds at the edge of these H\textsc{ii} regions, such as in the case of the H\textsc{ii} region associated with $\lambda$-Orionis \citep[e.g.,][]{Cepeda-Arroita2021}. Similarly, bright supernova remnants in the Galactic plane tend to show little evidence of AME \citep[e.g.,][]{Sun2011,Cruciani2016,Rennie2022}. Therefore, if we assume M\,31 is similar to our own Galaxy, then the total integrated emission will be dominated by these regions that have little or no AME component, and therefore reduce the average AME emissivity.

The apparent lack of AME in the integrated emission of other galaxies in the 10--50\,GHz range \citep{peel_2011,Bianchi2022} is also likely due to the same dilution over the entire volume of the galaxy. However, more observations are needed to determine whether this is true in general. This point is further supported by observations of several galaxies \citep{Murphy2012}, including NGC~6946 \citep{Hensley2015} and NGC~4725 \citep{Murphy2018}, which found AME within only a few extranuclear regions.

One reason why it is important to determine the contribution of AME to the integrated spectrum of galaxies in general is that free-free emission can be used to determine star-formation rates. It has been suggested that 30\,GHz is an ideal target frequency where the synchrotron and thermal dust contributions are minimal \citep{murphy2011}. If AME is also a significant contribution, then this will bias any SFR estimates to lower values.

\subsection{Star-Formation in M31}

Free-free emission is directly proportional to the star-formation rate of a galaxy \citep{murphy2011}. Using the updated values for fitted emission measure, and an assumed electron temperature of 8000\,K, we can estimate the star-formation rate of M\,31 using one of several derived scaling relations \citep{kennicutt1998,condon1998,murphy2011}. Here we adopt the relations given in \citet{murphy2011}. When fitting for spinning dust emission with a power-law synchrotron model we do not find a significant detection of the emission measure: $\mathrm{EM} =~$\CMBSRCsubEm\,cm$^{-6}$\,pc, thus we can only put a $1 \sigma$ upper limit on the expected star-formation rate of \amePlawMurphySFR. If we do not fit for AME, the EM is higher at $\mathrm{EM} =$\CMBSRCsubnoameEm\,cm$^{-6}$\,pc, which corresponds to a flux density of \noamePlawFFFlux\,Jy at 30\,GHz, and a star-formation rate of \noamePlawMurphySFR. If we use the model including AME and a curved synchrotron component we find an emission measure similar to that found when not including the AME in the synchrotron power-law model: $\mathrm{EM} =~$\CMBSRCsubcurvedEm\,cm$^{-6}$\,pc, or \ameCurvatureFFFlux\,Jy at 30\,GHz, corresponding to a star-formation rate of \ameCurvatureMurphySFR.

The \textit{Planck} analysis found a star-formation rate for M\,31 of $0.12$\,$M_\odot$\,yr$^{-1}$, which is consistent within $1\,\sigma$ with our estimates. Estimates of the star-formation rate from H$\alpha$ give $0.44$\,$M_\odot$\,yr$^{-1}$ \citep{azimlu_2011}, IR/UV comparisons estimate $0.25 \pm 0.05$\,$M_\odot$\,yr$^{-1}$ \citep{ford_2013}, and \citet{xu1996} estimated $0.36 \pm 0.14$\,$M_\odot$\,yr$^{-1}$ using IRAS data. All of which are several times higher than the star-formation rates we have predicted from the integrated free-free emission of M\,31.

The finding that all the estimates of the total star-formation rate of M\,31 using continuum free-free emission are lower than those found using other tracers is surprising. There have been multiple surveys of the integrated emission from other galaxies that show that, in general, free-free estimates of star-formation rates are consistent with other tracers \citep[e.g.,][]{murphy2011}, however, AME is never included in these models, and generally there is only one measurement around the AME peak. There are several possible explanations for why our estimates of the star-formation rates are lower than expected. First, clearly the choice of model for the SED is critical, where models including AME predicted a systematically lower free-free emission contribution. However, the AME amplitude is robust and is supported by the data. Possibly then this indicates that the synchrotron emission spectrum is steepening at high frequencies, but even then our star-formation rates are lower than those predicted by other tracers. 

Another possible reason for the differences in star-formation rates relates to the models used to calibrate the star-formation rates derived for the different tracers. The calibration of star-formation tracers depends on assumptions about the initial mass function and stellar evolution models \citep{kennicutt1998}, which when measuring the star-formation rate from the global integrated emission, as we have done, can be very different to those of individual local regions \citep{Calzetti2013}, as was done for H$\alpha$ survey of M\,31 \citep{azimlu_2011}. Furthermore, when probing individual regions versus the global signal, very different timescales are being measured with the global signal being dominated by star-formation over the period of 10--100\,Myr, while local regions will be sensitive to more recent star-formation events of the order 1--10\,Myr \citep{Calzetti2013}. 

A further factor to consider when comparing the star-formation rates derived from H$\alpha$ and UV to radio derived star-formation rates is the effect of dust. H$\alpha$ and UV tracers will suffer from attenuation due to dust along the line-of-sight. It is possible to correct for dust absorption but it becomes increasingly difficult, and with larger uncertainties, as the dust opacity becomes large. If the attenuation due to dust is over corrected, then this would bias the H$\alpha$ and UV estimates high. 

Ultimately, it is difficult, given the large uncertainties on all of the star-formation rates presented, to give a definitive answer to the global star-formation rate of M\,31 other than it is approximately $0.3$\,$M_\odot$\,yr$^{-1}$. Additionally, we cannot rule out that AME dominates free-free emission in the global integrated signal of M\,31. This is critical to determine as free-free emission at 30\,GHz is commonly used to determine star-formation rates in surveys such as COLDz \citep{Algera2022}, and if AME generally dominates over free-free at 30\,GHz for other galaxies then this would bias star-formation rate estimates high. To answer this question will require future observations of M\,31 at arcminute resolutions so that the AME and free-free emission components of individual star-forming regions can be measured.

\section{Conclusions}
\label{sec:conclusions}

We have presented a new measurement of the integrated spectrum of M\,31, using C-BASS data as an additional constraint on the low-frequency emission.  We find that the integrated spectrum of M\,31 at microwave frequencies has contributions from synchrotron, free-free, and AME. Similarly to \citet{planck_andromeda}, we find that AME is marginally detected with a $3 \sigma$ significance of \CMBSRCsubSpAAme\,Jy. Further, we find that the average AME emissivity integrated over all of M\,31 to be \emissivity. This is much lower than the typical value found for AME sources; since we are measuring the average AME emissivity, this implies that the AME must not be uniformly distributed throughout M\,31--this is similar to what is observed in our Galaxy and other galaxies where AME has been observed \citep[e.g.,][]{Hensley2015}. We have showed that there are significant differences between the low-frequency fluxes observed by low resolution surveys like \mbox{C-BASS} and QUIJOTE, and the high resolution observations by the SRT \citep{battistelli_andromeda} and Effelsberg  \citep{berkhuijsen2003} that we expect are due to filtering of large-scale emission in the data processing of the high resolution data. 

We attempted to fit the spectrum of M\,31 using both power-law and curved power-law models of synchrotron emission. Although there is no statistical preference for synchrotron spectral curvature, the curved synchrotron models resulted in a higher emission measure and star-formation rate estimates that are more consistent with other star-formation tracers, but were still a factor of two or more lower. Therefore, we cannot entirely rule out the possibility of the synchrotron spectrum steepening at higher frequencies. 

In conclusion, we find tentative evidence for AME within M\,31; however, the low average emissivity suggests that it is not uniformly distributed throughout the ISM and instead is localised to smaller sub-regions within the galaxy. Recent results using dedicated observations of M\,31 from the QUIJOTE MFI are consistent with our results when using the same aperture photometry \citep{FernandezTorreiro2023}. However, the main results of the QUIJOTE analysis used the smaller apertures that match those given in \citet{battistelli_andromeda}, and in this case they find an AME flux density that is three times higher than what we find, and subsequently a much fainter (and steeper---$\alpha = -0.99\pm0.21$) synchrotron contribution. This clearly shows the importance of both constraining the low frequency synchrotron emission with data like \mbox{C-BASS} and also the choice of aperture that is used. Ultimately, there remains some uncertainty regarding the nature of AME within M\,31 that will require future follow-up observations at frequencies around the peak of the AME spectrum (20--30\,GHz) with arcminute resolution that can resolve individual AME emitting regions. Such observations are already underway at both 22\,GHz with the SRT \citep{Fatigoni2021}, and at 30\,GHz with COMAP \citep{Cleary2022}.

\section*{Acknowledgements}

The \mbox{C-BASS} project (\url{http://cbass.web.ox.ac.uk}) is a collaboration between Oxford and Manchester Universities in the UK, the California Institute of Technology in the U.S.A., Rhodes University, UKZN and the South African Radio Observatory in South Africa, and the King Abdulaziz City for Science and Technology (KACST) in Saudi Arabia. It has been supported by the NSF awards AST-0607857, AST-1010024, AST-1212217, and AST-1616227, and NASA award NNX15AF06G, the University of Oxford, the Royal Society, STFC, and the other participating institutions. SEH, CD, and JPL acknowledge support from an STFC Consolidated Grant (ST/P000649/1). This research was also supported by the South African Radio Astronomy Observatory, which is a facility of the National Research Foundation, an agency of the Department of Science and Technology. We would like to thank Russ Keeney for technical help at OVRO.  We make use of the \textsc{HEALPix} package \citep{Gorski2005},  IDL astronomy library and Python \textsc{Astropy} \citep{astropy:2013, astropy:2018}, \textsc{Matplotlib} \citep{Hunter:2007}, \textsc{Numpy} \citep{harris2020array}, \textsc{Healpy} \citep{Zonca2019}, \textsc{corner} \citep{corner} and \textsc{Scipy} \citep{2020SciPy-NMeth} packages. This research has made use of the NASA/IPAC Extragalactic Database (NED) which is operated by the Jet Propulsion Laboratory, California Institute of Technology, under contract with the National Aeronautics and Space Administration.
This research has also made use of data from the OVRO 40-m monitoring program \citep{OVRO_2011}, supported by private funding from the California Institute of Technology and the Max Planck Institute for Radio Astronomy, and by NASA grants NNX08AW31G, NNX11A043G, and NNX14AQ89G and NSF grants AST-0808050 and AST- 1109911.
Some of the presented results are based on observations obtained with the QUIJOTE experiment (\url{http://research.iac.es/proyecto/quijote}).

\appendix
\section{Source Catalogues}
\label{sec:sources_catalogues}

The integrated flux density of extragalactic point sources near M\,31 are comparable to the total integrated flux of M\,31 itself at lower frequencies ($\lesssim 20$\,GHz) \citep{berkhuijsen2003}. Therefore modelling and separating these background sources is critical to correctly estimate the relative contributions of free-free, synchrotron, and spinning dust emission associated with M\,31. Also, flat spectrum sources (e.g., AGN) can turn over at frequencies of tens of gigahertz~\citep{rani2011} mimicking a spinning dust spectrum. Fainter source populations tend towards a uniform Gaussian distribution across the sky, and hence will provide a similar flux density in both the annulus and aperture used for aperture photometry, i.e., they form part of the quasi-uniform background when the number of sources is large. We are primarily concerned with the brightest sources that contribute within the vicinity of M\,31, which can strongly bias the photometry.

\begin{figure}
    \centering
    \includegraphics[width=0.49\textwidth]{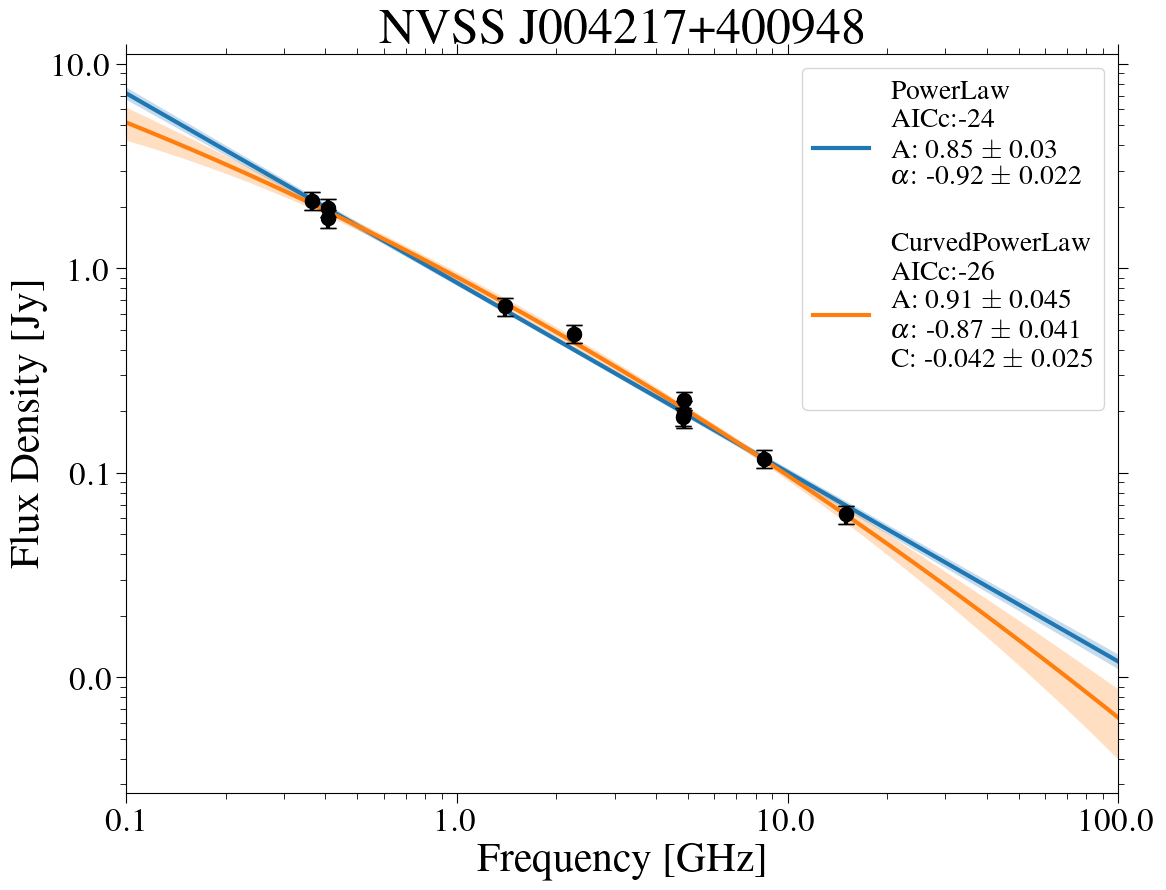}
        \includegraphics[width=0.49\textwidth]{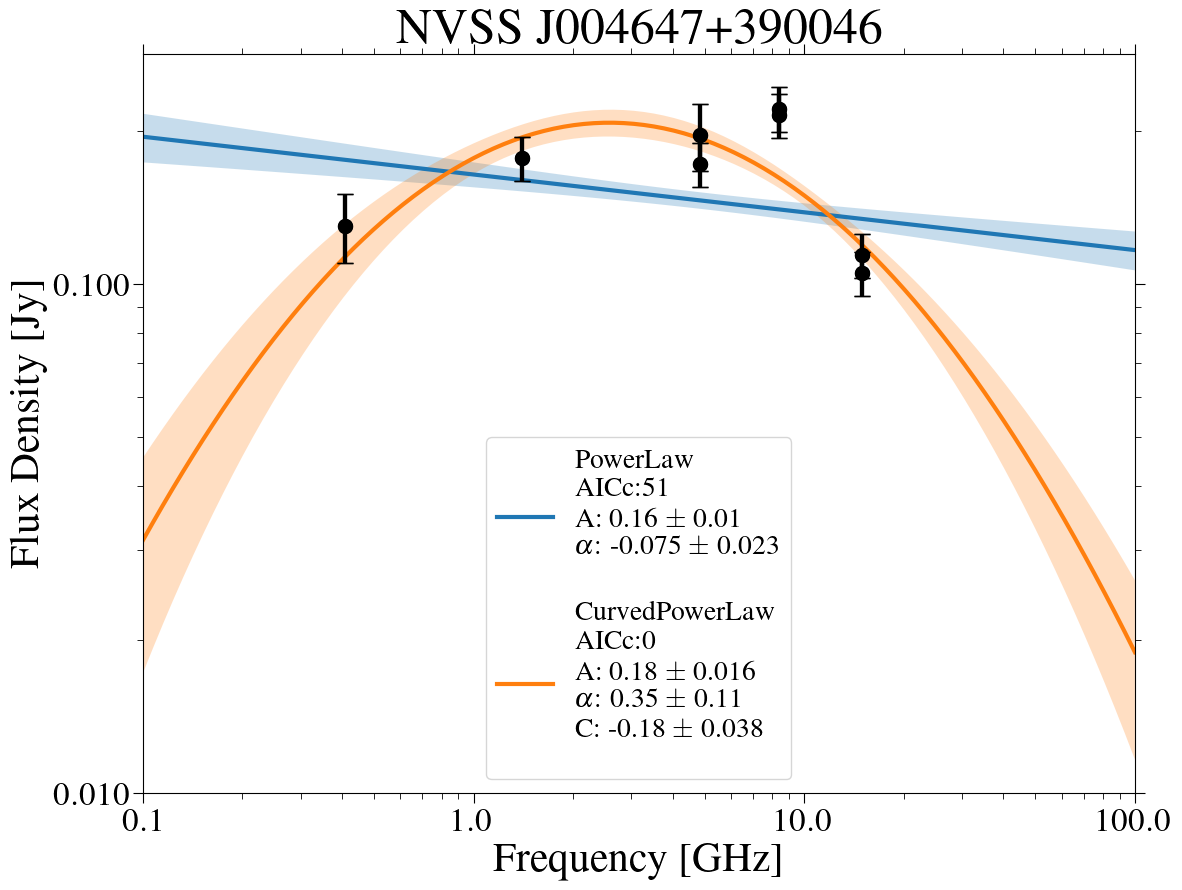}
    \caption{Best fit power-law and curved power-law models to the radio data for NVSS\,J004217+400948 (\textit{top}) and NVSS\,J004647+390046 (\textit{bottom}). The best fit parameters and AICc values for each model are shown in the legend.}
    \label{fig:point_sources_b}
\end{figure}

We used the  NRAO VLA Sky Survey \citep[NVSS][]{condon1998} catalogue to select bright sources at 1.4\,GHz within a region containing the aperture and background annulus centred on M\,31. We cross-referenced each source using the NASA/IPAC Extragalactic Database (NED)\footnote{https://ned.ipac.caltech.edu} to confirm each source was not associated with M\,31 itself. We used the historical photometry data available from NED to construct models for each source (\autoref{sec:source-models}) and then used these models to create point source maps at each frequency band that were then subtracted from the original data. All sources with only one frequency were discarded, and we also discarded one source with a rising spectrum, as the model predicted it would be extremely bright at microwave and sub-mm frequencies but this was not seen in any of the continuum surveys. After discarding bad sources, we were left with \sourcesfitted\, sources to construct our background source maps. Most of the background radio sources contribute less than a few percent of the total flux density from M\,31 except for the variable AGN \mbox{5C\,3.50} that has a flux density at 5 and 30\,GHz comparable to M\,31. A full summary of the modelled background radio sources is given in \autoref{tab:sources}. In \autoref{fig:point_sources} we show the fitted spectrum of \mbox{5C\,3.50}, which uses a combination of radio and infrared data to define the spectrum. For all other sources we fitted the models to just the available radio data. \autoref{fig:point_sources_b} shows examples of the fits to the second and third most bright radio sources near M\,31.

\begin{table*}
        \centering
        \caption{Modelled point sources within the photometry aperture around M\,31 in order of brightness at 5\,GHz. Sources are selected from the NVSS 1.4\,GHz survey \citep{condon1998} and cross-checked in the NED extragalactic source database. All source names use the NVSS designation except \mbox{5C\,3.50}, which is referenced in the main text. Sources that did not include ancillary data at more than one frequency or had steeply rising spectra are excluded. The source flux densities are from the model fits (see \autoref{sec:source-models} for details). }\label{tab:sources}
        \begin{tabular}{cccccccc}
        \hline
               &      &      &              & 408\,MHz     & 5\,GHz       & 30\,GHz      & \\ 
        Source & R.A. & Dec. & Angular Sep. & Flux Density & Flux Density & Flux Density & Model\\
         & (J2000) & (J2000) & (arcmin) & (Jy) & (mJy) & (mJy) &  \\
        \hline
        \hline
\mbox{5C\,3.50} (\mbox{B3 0035+413})  & $00^{\mathrm{h}}38^{\mathrm{m}}24.84^{\mathrm{s}}$ & $+41^\circ37{}^\prime06.00{}^{\prime\prime}$ & 52.91 & 604.8 $\pm$ 63.7 & 553.1 $\pm$ 56.5 & 518.9 $\pm$ 54.3 & Broken Power Law \\
NVSS J004217+400948 & $00^{\mathrm{h}}42^{\mathrm{m}}17.46^{\mathrm{s}}$ & $+40^\circ09{}^\prime48.5{}^{\prime\prime}$ & 66.50 & 1892.7 $\pm$ 98.2 & 200.4 $\pm$ 8.7 & 28.1 $\pm$ 4.7 & Curved Power Law \\
NVSS J004647+390046 & $00^{\mathrm{h}}46^{\mathrm{m}}47.52^{\mathrm{s}}$ & $+39^\circ00{}^\prime46.6{}^{\prime\prime}$ & 143.12 & 112.9 $\pm$ 18.1 & 191.6 $\pm$ 10.6 & 70.5 $\pm$ 10.1 & Curved Power Law \\
NVSS J004812+402152 & $00^{\mathrm{h}}48^{\mathrm{m}}12.94^{\mathrm{s}}$ & $+40^\circ21{}^\prime52.6{}^{\prime\prime}$ & 82.55 & 1516.1 $\pm$ 85.5 & 136.8 $\pm$ 8.6 & 24.5 $\pm$ 2.9 & Power Law \\
NVSS J004439+424801 & $00^{\mathrm{h}}44^{\mathrm{m}}39.17^{\mathrm{s}}$ & $+42^\circ48{}^\prime01.4{}^{\prime\prime}$ & 94.36 & 1325.0 $\pm$ 81.5 & 124.0 $\pm$ 4.7 & 22.8 $\pm$ 1.6 & Power Law \\
NVSS J003235+394215 & $00^{\mathrm{h}}32^{\mathrm{m}}35.51^{\mathrm{s}}$ & $+39^\circ42{}^\prime15.5{}^{\prime\prime}$ & 148.97 & 992.6 $\pm$ 49.8 & 118.7 $\pm$ 9.0 & 26.0 $\pm$ 3.6 & Power Law \\
NVSS J004218+412926 & $00^{\mathrm{h}}42^{\mathrm{m}}18.71^{\mathrm{s}}$ & $+41^\circ29{}^\prime26.8{}^{\prime\prime}$ & 14.15 & 1026.9 $\pm$ 67.7 & 113.2 $\pm$ 4.3 & 23.4 $\pm$ 1.7 & Power Law \\
NVSS J004802+393726 & $00^{\mathrm{h}}48^{\mathrm{m}}02.06^{\mathrm{s}}$ & $+39^\circ37{}^\prime26.6{}^{\prime\prime}$ & 115.75 & 220.3 $\pm$ 14.5 & 111.6 $\pm$ 8.2 & 68.7 $\pm$ 9.7 & Power Law \\
NVSS J003506+423818 & $00^{\mathrm{h}}35^{\mathrm{m}}06.02^{\mathrm{s}}$ & $+42^\circ38{}^\prime18.4{}^{\prime\prime}$ & 118.34 & 923.9 $\pm$ 64.3 & 110.0 $\pm$ 9.7 & 24.0 $\pm$ 3.9 & Power Law \\
NVSS J005331+402721 & $00^{\mathrm{h}}53^{\mathrm{m}}31.76^{\mathrm{s}}$ & $+40^\circ27{}^\prime21.5{}^{\prime\prime}$ & 131.81 & 666.0 $\pm$ 45.7 & 88.3 $\pm$ 7.1 & 20.8 $\pm$ 3.0 & Power Law \\
NVSS J005413+384214 & $00^{\mathrm{h}}54^{\mathrm{m}}13.78^{\mathrm{s}}$ & $+38^\circ42{}^\prime14.5{}^{\prime\prime}$ & 202.79 & 188.7 $\pm$ 45.2 & 82.4 $\pm$ 4.7 & 45.5 $\pm$ 10.2 & Power Law \\
NVSS J005832+390209 & $00^{\mathrm{h}}58^{\mathrm{m}}32.28^{\mathrm{s}}$ & $+39^\circ02{}^\prime09.2{}^{\prime\prime}$ & 225.30 & 310.7 $\pm$ 82.3 & 81.7 $\pm$ 13.6 & 31.4 $\pm$ 13.7 & Power Law \\
NVSS J005455+382150 & $00^{\mathrm{h}}54^{\mathrm{m}}55.87^{\mathrm{s}}$ & $+38^\circ21{}^\prime50.9{}^{\prime\prime}$ & 223.84 & 1261.7 $\pm$ 65.8 & 77.1 $\pm$ 4.9 & 10.5 $\pm$ 1.2 & Power Law \\
NVSS J003035+421128 & $00^{\mathrm{h}}30^{\mathrm{m}}35.15^{\mathrm{s}}$ & $+42^\circ11{}^\prime28{}^{\prime\prime}$ & 146.81 & 114.9 $\pm$ 75.6 & 69.4 $\pm$ 10.9 & 48.5 $\pm$ 18.6 & Power Law \\
NVSS J003216+400711 & $00^{\mathrm{h}}32^{\mathrm{m}}16.24^{\mathrm{s}}$ & $+40^\circ07{}^\prime11.2{}^{\prime\prime}$ & 137.51 & 285.1 $\pm$ 19.8 & 66.0 $\pm$ 4.9 & 23.2 $\pm$ 3.3 & Power Law \\
NVSS J004154+392521 & $00^{\mathrm{h}}41^{\mathrm{m}}54.96^{\mathrm{s}}$ & $+39^\circ25{}^\prime21{}^{\prime\prime}$ & 111.16 & 1026.5 $\pm$ 50.4 & 65.5 $\pm$ 2.4 & 9.2 $\pm$ 0.6 & Power Law \\
NVSS J005332+403443 & $00^{\mathrm{h}}53^{\mathrm{m}}32.35^{\mathrm{s}}$ & $+40^\circ34{}^\prime43.6{}^{\prime\prime}$ & 129.27 & 420.0 $\pm$ 37.1 & 63.5 $\pm$ 10.7 & 16.5 $\pm$ 4.9 & Power Law \\
NVSS J004843+404458 & $00^{\mathrm{h}}48^{\mathrm{m}}43.83^{\mathrm{s}}$ & $+40^\circ44{}^\prime58.5{}^{\prime\prime}$ & 74.68 & 358.7 $\pm$ 23.6 & 62.1 $\pm$ 5.0 & 17.7 $\pm$ 2.7 & Power Law \\
NVSS J004654+435328 & $00^{\mathrm{h}}46^{\mathrm{m}}54.58^{\mathrm{s}}$ & $+43^\circ53{}^\prime28.3{}^{\prime\prime}$ & 163.97 & 410.5 $\pm$ 31.7 & 58.7 $\pm$ 4.6 & 14.6 $\pm$ 2.2 & Power Law \\
NVSS J003638+425229 & $00^{\mathrm{h}}36^{\mathrm{m}}38.62^{\mathrm{s}}$ & $+42^\circ52{}^\prime29.2{}^{\prime\prime}$ & 117.83 & 1196.9 $\pm$ 70.6 & 57.9 $\pm$ 4.2 & 6.6 $\pm$ 0.9 & Power Law \\
NVSS J004755+394900 & $00^{\mathrm{h}}47^{\mathrm{m}}55.49^{\mathrm{s}}$ & $+39^\circ49{}^\prime00.2{}^{\prime\prime}$ & 105.31 & 180.0 $\pm$ 19.9 & 56.2 $\pm$ 9.2 & 24.4 $\pm$ 7.2 & Power Law \\
NVSS J004441+382957 & $00^{\mathrm{h}}44^{\mathrm{m}}41.56^{\mathrm{s}}$ & $+38^\circ29{}^\prime57.9{}^{\prime\prime}$ & 167.67 & 135.2 $\pm$ 17.2 & 55.8 $\pm$ 2.8 & 29.7 $\pm$ 3.7 & Power Law \\
NVSS J003111+394156 & $00^{\mathrm{h}}31^{\mathrm{m}}11.43^{\mathrm{s}}$ & $+39^\circ41{}^\prime56.1{}^{\prime\prime}$ & 161.89 & 439.3 $\pm$ 26.3 & 55.6 $\pm$ 6.3 & 12.7 $\pm$ 2.6 & Power Law \\
NVSS J003018+380355 & $00^{\mathrm{h}}30^{\mathrm{m}}18.81^{\mathrm{s}}$ & $+38^\circ03{}^\prime55.4{}^{\prime\prime}$ & 239.76 & 227.9 $\pm$ 18.6 & 53.2 $\pm$ 3.7 & 18.8 $\pm$ 2.6 & Power Law \\
NVSS J003525+394047 & $00^{\mathrm{h}}35^{\mathrm{m}}25.6^{\mathrm{s}}$ & $+39^\circ40{}^\prime47.7{}^{\prime\prime}$ & 126.63 & 330.8 $\pm$ 22.1 & 53.0 $\pm$ 4.0 & 14.3 $\pm$ 1.9 & Power Law \\
NVSS J004354+404634 & $00^{\mathrm{h}}43^{\mathrm{m}}54.35^{\mathrm{s}}$ & $+40^\circ46{}^\prime34.2{}^{\prime\prime}$ & 32.39 & 421.0 $\pm$ 30.9 & 52.5 $\pm$ 4.8 & 11.9 $\pm$ 2.1 & Power Law \\
NVSS J003022+381723 & $00^{\mathrm{h}}30^{\mathrm{m}}22.53^{\mathrm{s}}$ & $+38^\circ17{}^\prime23.2{}^{\prime\prime}$ & 228.52 & 634.9 $\pm$ 36.2 & 52.3 $\pm$ 5.9 & 8.8 $\pm$ 1.8 & Power Law \\
NVSS J004601+435517 & $00^{\mathrm{h}}46^{\mathrm{m}}01.85^{\mathrm{s}}$ & $+43^\circ55{}^\prime17.4{}^{\prime\prime}$ & 163.28 & 819.8 $\pm$ 63.1 & 51.9 $\pm$ 8.8 & 7.2 $\pm$ 2.2 & Power Law \\
NVSS J003419+424659 & $00^{\mathrm{h}}34^{\mathrm{m}}19.26^{\mathrm{s}}$ & $+42^\circ46{}^\prime59.4{}^{\prime\prime}$ & 130.55 & 593.3 $\pm$ 45.3 & 51.7 $\pm$ 4.6 & 9.0 $\pm$ 1.5 & Power Law \\
NVSS J005405+412515 & $00^{\mathrm{h}}54^{\mathrm{m}}05.78^{\mathrm{s}}$ & $+41^\circ25{}^\prime15.2{}^{\prime\prime}$ & 128.28 & 170.0 $\pm$ 17.6 & 51.2 $\pm$ 7.8 & 21.7 $\pm$ 5.7 & Power Law \\
NVSS J005744+415410 & $00^{\mathrm{h}}57^{\mathrm{m}}44.68^{\mathrm{s}}$ & $+41^\circ54{}^\prime10.8{}^{\prime\prime}$ & 172.65 & 404.2 $\pm$ 25.3 & 51.1 $\pm$ 5.8 & 11.6 $\pm$ 2.3 & Power Law \\
NVSS J004349+383010 & $00^{\mathrm{h}}43^{\mathrm{m}}49.42^{\mathrm{s}}$ & $+38^\circ30{}^\prime10.1{}^{\prime\prime}$ & 166.42 & 684.4 $\pm$ 38.3 & 49.3 $\pm$ 6.1 & 7.5 $\pm$ 1.7 & Power Law \\
NVSS J003137+391904 & $00^{\mathrm{h}}31^{\mathrm{m}}37.9^{\mathrm{s}}$ & $+39^\circ19{}^\prime04.3{}^{\prime\prime}$ & 172.70 & 472.0 $\pm$ 34.7 & 49.0 $\pm$ 6.5 & 9.7 $\pm$ 2.4 & Power Law \\
NVSS J003406+430242 & $00^{\mathrm{h}}34^{\mathrm{m}}06.03^{\mathrm{s}}$ & $+43^\circ02{}^\prime42.8{}^{\prime\prime}$ & 143.44 & 410.1 $\pm$ 29.0 & 48.8 $\pm$ 5.4 & 10.7 $\pm$ 2.1 & Power Law \\
NVSS J004358+393756 & $00^{\mathrm{h}}43^{\mathrm{m}}58.38^{\mathrm{s}}$ & $+39^\circ37{}^\prime56.6{}^{\prime\prime}$ & 99.19 & 378.4 $\pm$ 23.0 & 48.6 $\pm$ 3.6 & 11.2 $\pm$ 1.5 & Power Law \\
NVSS J005044+381310 & $00^{\mathrm{h}}50^{\mathrm{m}}44.93^{\mathrm{s}}$ & $+38^\circ13{}^\prime10.8{}^{\prime\prime}$ & 204.95 & 423.6 $\pm$ 30.2 & 48.4 $\pm$ 2.9 & 10.3 $\pm$ 1.3 & Power Law \\
NVSS J003428+403556 & $00^{\mathrm{h}}34^{\mathrm{m}}28.73^{\mathrm{s}}$ & $+40^\circ35{}^\prime56.4{}^{\prime\prime}$ & 101.80 & 385.7 $\pm$ 26.8 & 48.4 $\pm$ 6.2 & 11.0 $\pm$ 2.6 & Power Law \\
NVSS J003007+433524 & $00^{\mathrm{h}}30^{\mathrm{m}}07.93^{\mathrm{s}}$ & $+43^\circ35{}^\prime24.2{}^{\prime\prime}$ & 197.11 & 220.3 $\pm$ 16.2 & 45.0 $\pm$ 5.1 & 14.5 $\pm$ 2.9 & Power Law \\
NVSS J003641+380903 & $00^{\mathrm{h}}36^{\mathrm{m}}41.28^{\mathrm{s}}$ & $+38^\circ09{}^\prime03.8{}^{\prime\prime}$ & 199.63 & 675.3 $\pm$ 33.3 & 44.0 $\pm$ 6.1 & 6.2 $\pm$ 1.5 & Power Law \\
NVSS J003039+423704 & $00^{\mathrm{h}}30^{\mathrm{m}}39.67^{\mathrm{s}}$ & $+42^\circ37{}^\prime04.9{}^{\prime\prime}$ & 157.14 & 309.9 $\pm$ 24.8 & 43.8 $\pm$ 7.3 & 10.8 $\pm$ 3.1 & Power Law \\
NVSS J004824+431931 & $00^{\mathrm{h}}48^{\mathrm{m}}24.46^{\mathrm{s}}$ & $+43^\circ19{}^\prime31.2{}^{\prime\prime}$ & 138.53 & 209.2 $\pm$ 17.8 & 43.3 $\pm$ 3.0 & 14.0 $\pm$ 2.0 & Power Law \\
NVSS J003048+411053 & $00^{\mathrm{h}}30^{\mathrm{m}}48.78^{\mathrm{s}}$ & $+41^\circ10{}^\prime53.7{}^{\prime\prime}$ & 134.58 & 1232.7 $\pm$ 73.3 & 43.3 $\pm$ 2.7 & 3.9 $\pm$ 0.4 & Power Law \\
NVSS J005213+424102 & $00^{\mathrm{h}}52^{\mathrm{m}}13.75^{\mathrm{s}}$ & $+42^\circ41{}^\prime02.9{}^{\prime\prime}$ & 135.73 & 462.9 $\pm$ 30.3 & 42.9 $\pm$ 5.6 & 7.8 $\pm$ 1.8 & Power Law \\
NVSS J003736+393812 & $00^{\mathrm{h}}37^{\mathrm{m}}36.72^{\mathrm{s}}$ & $+39^\circ38{}^\prime12.4{}^{\prime\prime}$ & 114.03 & 192.8 $\pm$ 21.0 & 42.1 $\pm$ 1.8 & 0.3 $\pm$ 0.1 & Curved Power Law \\
NVSS J004844+395331 & $00^{\mathrm{h}}48^{\mathrm{m}}44.72^{\mathrm{s}}$ & $+39^\circ53{}^\prime31.5{}^{\prime\prime}$ & 107.29 & 304.5 $\pm$ 21.2 & 41.7 $\pm$ 3.1 & 10.0 $\pm$ 1.4 & Power Law \\
NVSS J005537+383212 & $00^{\mathrm{h}}55^{\mathrm{m}}37.54^{\mathrm{s}}$ & $+38^\circ32{}^\prime12.2{}^{\prime\prime}$ & 221.04 & 264.9 $\pm$ 18.9 & 41.6 $\pm$ 4.7 & 11.1 $\pm$ 2.2 & Power Law \\
NVSS J003719+385916 & $00^{\mathrm{h}}37^{\mathrm{m}}19.29^{\mathrm{s}}$ & $+38^\circ59{}^\prime16.5{}^{\prime\prime}$ & 150.25 & 590.0 $\pm$ 43.4 & 41.1 $\pm$ 6.0 & 6.1 $\pm$ 1.6 & Power Law \\
NVSS J003848+411607 & $00^{\mathrm{h}}38^{\mathrm{m}}48.28^{\mathrm{s}}$ & $+41^\circ16{}^\prime07.4{}^{\prime\prime}$ & 44.29 & 170.0 $\pm$ 19.4 & 39.0 $\pm$ 3.0 & 13.6 $\pm$ 2.0 & Power Law \\
NVSS J005411+421634 & $00^{\mathrm{h}}54^{\mathrm{m}}11.47^{\mathrm{s}}$ & $+42^\circ16{}^\prime34.8{}^{\prime\prime}$ & 141.71 & 463.1 $\pm$ 29.4 & 36.2 $\pm$ 3.3 & 5.9 $\pm$ 0.9 & Power Law \\
NVSS J003956+411138 & $00^{\mathrm{h}}39^{\mathrm{m}}56.35^{\mathrm{s}}$ & $+41^\circ11{}^\prime38.4{}^{\prime\prime}$ & 31.84 & 187.1 $\pm$ 14.2 & 36.2 $\pm$ 6.0 & 11.2 $\pm$ 3.2 & Power Law \\
NVSS J005351+424936 & $00^{\mathrm{h}}53^{\mathrm{m}}51.6^{\mathrm{s}}$ & $+42^\circ49{}^\prime36.1{}^{\prime\prime}$ & 155.23 & 120.0 $\pm$ 18.9 & 35.5 $\pm$ 5.9 & 14.8 $\pm$ 4.8 & Power Law \\
NVSS J005540+393337 & $00^{\mathrm{h}}55^{\mathrm{m}}40.34^{\mathrm{s}}$ & $+39^\circ33{}^\prime37.8{}^{\prime\prime}$ & 179.81 & 469.2 $\pm$ 35.2 & 35.3 $\pm$ 5.6 & 5.5 $\pm$ 1.6 & Power Law \\
NVSS J004942+382259 & $00^{\mathrm{h}}49^{\mathrm{m}}42.41^{\mathrm{s}}$ & $+38^\circ22{}^\prime59.5{}^{\prime\prime}$ & 190.84 & 226.2 $\pm$ 15.7 & 35.3 $\pm$ 6.2 & 9.3 $\pm$ 2.8 & Power Law \\
NVSS J005755+424606 & $00^{\mathrm{h}}57^{\mathrm{m}}55.38^{\mathrm{s}}$ & $+42^\circ46{}^\prime06.1{}^{\prime\prime}$ & 191.68 & 200.0 $\pm$ 19.9 & 35.2 $\pm$ 5.7 & 10.2 $\pm$ 2.9 & Power Law \\
NVSS J005057+420452 & $00^{\mathrm{h}}50^{\mathrm{m}}57.43^{\mathrm{s}}$ & $+42^\circ04{}^\prime52.1{}^{\prime\prime}$ & 104.24 & 242.8 $\pm$ 21.0 & 35.0 $\pm$ 5.5 & 8.7 $\pm$ 2.5 & Power Law \\
NVSS J004653+394254 & $00^{\mathrm{h}}46^{\mathrm{m}}53.45^{\mathrm{s}}$ & $+39^\circ42{}^\prime54.4{}^{\prime\prime}$ & 104.58 & 130.0 $\pm$ 21.6 & 34.4 $\pm$ 4.9 & 13.3 $\pm$ 3.7 & Power Law \\
NVSS J005110+383842 & $00^{\mathrm{h}}51^{\mathrm{m}}10.93^{\mathrm{s}}$ & $+38^\circ38{}^\prime42.5{}^{\prime\prime}$ & 184.96 & 140.0 $\pm$ 19.0 & 34.4 $\pm$ 4.5 & 12.6 $\pm$ 3.0 & Power Law \\
NVSS J004648+420855 & $00^{\mathrm{h}}46^{\mathrm{m}}48.11^{\mathrm{s}}$ & $+42^\circ08{}^\prime55.5{}^{\prime\prime}$ & 69.74 & 400.7 $\pm$ 23.5 & 33.9 $\pm$ 4.8 & 5.8 $\pm$ 1.4 & Power Law \\
\hline
\end{tabular}

\end{table*}

\section*{Data Availability}

All model fits for M\,31 and radio sources can be made available upon request. Ancillary datasets described in \autoref{sec:ancillary} are available via from the papers cited, the NASA LAMBDA website, or the \textit{Planck} legacy archive. The C-BASS map is not currently available but will be published (Taylor et al., in prep.) and released in the near future.



\bibliographystyle{mnras}
\bibliography{example} 




\appendix


\bsp	
\label{lastpage}
\end{document}